\documentclass{pasj00}  % PASJ
\draft  % PASJ

\newcommand{\vvec}{{\mathbf v}} %_\perp}
\newcommand{\bvec}{{\mathbf B}}
\newcommand{\Svec}{{\mathbf S}} 
\newcommand{\zvec}{{\mathbf z}}
\newcommand{\evec}{{\mathbf E}}
\newcommand{\be}{\begin{equation}}
\newcommand{\ee}{\end{equation}}
\newcommand{\hatz}{\hat \zvec}
 
\newcommand{\bnfi}{B_{\rm NFI}}

\def\lapprox{\mathrel{\hbox{\rlap{\hbox{\lower4pt\hbox{$\sim$}}}\hbox{$<$}}}}
\def\gapprox{\mathrel{\hbox{\rlap{\hbox{\lower4pt\hbox{$\sim$}}}\hbox{$>$}}}}

\begin{document}
\SetRunningHead{Welsch}{Photospheric Poynting Flux} % PASJ

\Received{2014/02/28}%{yyyy/mm/dd}   (could instead be Feb. 18, for v.1)
\Accepted{2014/12/02}%{yyyy/mm/dd}

\title{The Photospheric Poynting Flux and Coronal Heating}

\author{B. T. \textsc{Welsch}}
\affil{Space Sciences Laboratory, UC, Berkeley}
\email{welsch@ssl.berkeley.edu}


\maketitle

\begin{abstract}
Some models of coronal heating suppose that convective motions at the
photosphere shuffle the footpoints of coronal magnetic fields and
thereby inject sufficient magnetic energy upward to account for
observed coronal and chromospheric energy losses in active regions.
Using high-resolution observations of plage magnetic fields made with
the Solar Optical Telescope aboard the {\em Hinode} satellite, we
investigate this idea by estimating the upward transport of magnetic
energy --- the vertical Poynting flux, $S_z$ --- across the
photosphere in a plage region.  To do so, we combine: (i) estimates of
photospheric horizontal velocities, $\vvec_h$, determined by local
correlation tracking applied to a sequence of line-of-sight magnetic
field maps from the Narrowband Filter Imager, with (ii) a vector
magnetic field measurement from the SpectroPolarimeter.  Plage fields
are ideal observational targets for estimating energy injection by
convection, because they are: (i) strong enough to be measured with
relatively small uncertainties; (ii) not so strong that convection is
heavily suppressed (as within umbrae); and (iii) unipolar, so $S_z$ in
plage is not influenced by mixed-polarity processes (e.g., flux
emergence) unrelated to heating in stable, active-region fields. In
this plage region, we found that the average $S_z$ varied in space,
but was positive (upward) and sufficient to explain coronal heating,
with values near $(5 \pm 1) \times 10^7$ erg cm$^{-2}$ s$^{-1}$. We
find the energy input per unit magnetic flux to be on the order of
$10^5$ erg s$^{-1}$ Mx$^{-1}$.  A comparison of intensity in a Ca II
image co-registered with one plage magnetogram shows stronger spatial
correlations with both total field strength and unsigned vertical
field, $|B_z|,$ than either $S_z$ or horizontal flux density, $B_h$.
The observed Ca II brightness enhancement, however, probably contains
a strong contribution from a near-photosphere hot-wall effect, which
is unrelated to heating in the solar atmosphere.
\end{abstract}
%

%Figure~\ref{fig:usage} is an example of the `figure' environment.
%\begin{figure}
% \begin{center}
%  \includegraphics[width=8cm]{fig1_maybe_old.eps}
% \end{center}
% \caption{A figure.}\label{fig:usage}
%\end{figure}

\section{Introduction}

How is the solar corona heated to temperatures of $\sim 1$ MK, when
the lower layers of the solar atmosphere are $\sim 10^4$ K or less?
Evidently, the energy needed to heat the Sun's atmosphere must cross
the photosphere in some organized form before being converted into
heat (disorganized, ``thermalized'' energy) in the chromosphere and
corona.  Because the magnetic fields that permeate the corona are all
anchored at the photosphere, they are natural candidates for energetic
coupling between the solar interior and corona.  In the interior,
motions in the Sun's gas are driven by convection, and some fraction
of the kinetic energy in turbulent convective motions is thought to be
converted into energy stored in electric currents flowing in coronal
magnetic fields that is then dissipated as heat.
These induced currents might be characterized as either steady or
rapidly varying (e.g., wave-driven) relative to the timescales of the
atmospheric response, and the dissipation of each has been referred to
as ``DC'' (direct-current) or ``AC'' (alternating-current) heating,
respectively (e.g., Klimchuk 2006). \nocite{Klimchuk2006}

To be a viable coronal heating mechanism, the input energy must be
commensurate with observed energy losses in active region (AR) fields,
estimated by \citet{Withbroe1977} to be $\sim 1 \times 10^7$ erg
cm$^{-2}$ s$^{-1}$ for the corona and $\sim 2 \times 10^7$ erg
cm$^{-2}$ s$^{-1}$ for the chromosphere.  Waves were once thought to
be primarily responsible for coronal and chromospheric heating (see,
e.g., Withbroe and Noyes 1977).  \nocite{Withbroe1977} While waves
(e.g., Tomczyk \etal \,2007) \nocite{Tomczyk2007} and wave dissipation
(e.g., Hahn \etal \,2012) \nocite{Hahn2012} have been reported in the
corona, currently available observations have not demonstrated that
they supply sufficient energy to heat the active-region chromosphere,
transition region, and corona.  In contrast to models invoking
dissipation of waves, other models posit that convective motions
induce relatively long-lived, DC currents that are episodically
dissipated to heat the chromosphere and corona.  We explore the latter
hypothesis here.

There is a long history of modeling this convection-driven coronal
energy input.  Parker (1983a, 1983b) \nocite{Parker1983, Parker1983b}
proposed that convection braids and twists the photospheric footpoints
of coronal magnetic fields, and thereby injects energy into the
corona.  This energy is stored in current sheets, and is transiently
dissipated in small bursts referred to as nanoflares
\citep{Parker1988}, with typical energies of $\sim 10^{24}$ erg, about
$10^{-9}$ of the energies in very large flares.
\citet{Galsgaard1996} modeled an idealization of this process by
imposing shearing flows on the upper and lower boundaries of an
initially uniform field in an MHD simulation, and found sufficient
power to heat the corona. \citet{Gudiksen2002} imposed a more complex
flow field, meant to mimic convective motions, on an MHD model of the
coronal field and also found sufficient power, as well as morphology
consistent with aspects of coronal observations.  In the framework of
reduced MHD, \citet{Rappazzo2008}, also found sufficient power, even
though fields in their model were only weakly braided.
More recently, \citet{Bingert2011} also modeled this process in MHD
with a detailed treatment of the energy equation and found heating
that is transient in time and space, and concentrated in and near the
modeled transition region.

One promising observational approach to constraining models of coronal
heating is to analyze time evolution of magnetic fields at the
photosphere, where the magnetic field is precisely and routinely
measured.  Clear evidence of braiding or twisting motions would
support the mechanism proposed by Parker.  \citet{Schrijver1998}
proposed that continuous emergence and cancellation of small-scale
fields in the quiet Sun's ``magnetic carpet'' leads to reconnection
and heating, but \citet{Close2004} used sequential potential models of
quiet-Sun fields to argue that emergence and cancellation are not
required: just reconnection between existing flux systems, as their
photospheric footpoints move, should be sufficient.  \citet{Meyer2013}
recently directly incorporated magnetogram sequences into the lower
boundary of a magnetofrictional model of quiet-sun coronal field
evolution, to investigate the dissipation of magnetic energy within
the simulation.  Aspects of energy dissipation in their model were
qualitatively consistent with solar observations, although their total
upward energy flux was smaller than the observationally estimated
energy demand for the quiet-sun atmosphere.

\citet{Yeates2014} recently investigated analytic expressions for
lower bounds on the upward-directed Poynting flux of magnetic energy
in a region of plage fields in NOAA AR 10930, based upon observed
photospheric magnetic and velocity fields.  The flows they analyzed
were estimated by \citet{Welsch2012}, who applied Fourier Local
Correlation Tracking (FLCT; Fisher and Welsch 2008)
\nocite{Fisher2008} to a sequence of line-of-sight magnetograms
(magnetic field maps) of this active region.  These 
%``line-wing''
magnetograms were observed with the Narrowband Filter Imager (NFI)
instrument on the Solar Optical Telescope (SOT) \citep{Tsuneta2008,
  Suematsu2008, Ichimoto2008, Shimizu2008} aboard the {\em
  Hinode} satellite \citep{Kosugi2007}, with a cadence $\simeq 120$ s,
over about 13 hours on 2006 December 12 -- 13.
%broadband filter imager (BFI) of the SOT \citep{Tsuneta2008} aboard
%the {\em Hinode} satellite \citep{Kosugi2007}. 

\citet{Yeates2014} compared their lower bounds on the Poynting flux
with a direct estimate of the Poynting flux, obtained with a procedure
that we explain in detail here. First, they assumed that the photospheric
magnetic field, $\bvec$, is frozen to the plasma --- a valid
assumption in quite general circumstances (see, e.g., Parker 1984).
\nocite{Parker1984a} Then the photospheric electric field, $\evec$,
is ideal, and equal to $-(\vvec \times \bvec)/c$, where $\vvec$ is
the photospheric velocity.  Then the (vector) Poynting flux of
magnetic energy, $\Svec,$ can be expressed in terms of $\vvec$ and
$\bvec$ as
\be \Svec = \bvec \times (\vvec \times \bvec)/4 \pi 
~. \label{eqn:poynting_vec} \ee
Approximating the photospheric surface as locally planar, we adopt
Cartesian geometry, and use $z$ and $h$ to refer to vertical and
horizontal directions, respectively.  Then the vertical component of
the Poynting flux is
\be S_z = [v_z B_h^2 - (\vvec_h \cdot \bvec_h)B_z]/4\pi
~. \label{eqn:poynting_vert} \ee
This expression for total Poynting flux has been conceptually divided
into an ``emergence'' term, which contains $v_z$, and a ``shear''
term, which contains $\vvec_h$ \citep{Liu2012d, Parnell2012}. 

We digress for a moment to note that a positive (upward) value for the
shearing term also implies the emergence of magnetized plasma across
the photosphere.
A clear example of this is the special case in which $v_z$ is zero (so
the emergence term vanishes), and the shearing term is positive.
Then both $B_z$ and $\bvec_h$ must be nonzero, implying $\bvec$ is
tilted; and $\vvec_h$ must have a nonzero projection onto $\bvec_h$.
We define the component of $\vvec_h$ along the horizontal field
$\bvec_h$ to be $\vvec_{Bh}$, and then further decompose $\vvec_{Bh}$ into a
component parallel to the total field $\bvec$, which we label
$v_{Bh,\parallel}$, and a component perpendicular to $\bvec$, which we
label $v_{Bh,\perp}$.
Since $\bvec$ is tilted, $v_{Bh,\perp}$ must also be; and it 
must be tilted upward when the shearing term is positive.
This upward tilt for $v_{Bh,\perp}$ implies that this component of the
flow advects the tilted magnetic flux upward.
%
%(D\'emouin and Berger [2003] \nocite{Demoulin2003} discuss the
%Poynting flux in terms of photospheric flows, and their Figure 1. is
%relevant here: their $\uvec_f$ plays the role of $\vvec_{Bh}$ here,
%and clearly the component of their $\uvec_f$ perpendicular to their
%$\bvec$ has a vertical [upward] component.)
%
The parallel flow $v_{Bh,\parallel}$ neither advects magnetic field or
produces a Poynting flux.
The counter-intuitive result that a horizontal velocity can produce
upward transport of magnetic fields arises because the total velocity
included a component of the velocity parallel to $\bvec$, which is
irrelevant for the Poynting flux.
(In the special case that the parallel velocity is zero, then equation
[\ref{eqn:poynting_vec}] reduces to $\Svec = B^2 \vvec/4\pi$, and 
equation [\ref{eqn:poynting_vec}] becomes $S_z = B^2 v_z/4\pi$.) 
%
%(We ignore flows parallel to $\bvec$ here, which cancel out in
%equation \ref{eqn:poynting_vert}.)  Such upward transport of a tilted
%magnetic field does not increase the total unsigned magnetic flux
%threading the photosphere, in contrast to the emergence of new (or
%additional) magnetic flux, which must occur at a polarity inversion
%line.
%

Since \citet{Yeates2014} were primarily focused on
heating in plage --- regions of nearly-vertical field when new flux is not
emerging --- the shearing term should dominate, meaning
\be S_z^{\rm \,plage} \simeq - (\vvec_h \cdot \bvec_h)B_z/4\pi
~. \label{eqn:poynting_shear} \ee
%  (\ref{eqn:poynting_shear})
%

\citet{Yeates2014} treated the flows estimated by FLCT as horizontal
velocities.  We note that there is some controversy about how to
interpret of velocities determined by correlation tracking and other
``optical flow'' \citep{Schuck2006} methods.  \citet{Demoulin2003}
suggested that the apparently horizontal flows estimated by LCT
%, which we denote $\uvec$, 
are a linear combination of the horizontal velocity with the vertical
velocity, with weighting determined by the ratio of horizontal to
vertical magnetic field.  
To test the accuracy of velocities reconstructed from magnetogram
sequences, \citet{Welsch2007} compared flows estimated by several
methods, including LCT, using synthetic magnetograms extracted from
MHD simulations of an emerging magnetic flux tube in the solar
interior in which the actual velocities were known.
Using the same test data, \citet{Schuck2008} subsequently argued that
optical flow methods, such as LCT, essentially estimate the horizontal
velocity, $\vvec_h$, although their estimates can be affected by
vertical flows.

The NFI magnetograms only provide estimates of the line-of-sight (LOS)
field, $B_{\rm LOS}$, but the expressions for the Poynting flux given
above all require knowledge of the vector magnetic field, $\bvec$.
Accordingly, \citet{Yeates2014} co-registered the (12 Mm $\times$ 12
Mm) region of the NFI field of view (FOV) that they studied with the
corresponding sub-region of a vector magnetogram observed by SOT's
SpectroPolarimeter 
%(SP; Tsuneta \etal \, 2008). \nocite{Tsuneta2008}
(SP; Lites \etal \, 2013). \nocite{Lites2013}
A reprojected vector magnetogram based upon these observations was
prepared by \citet{Schrijver2008} and is available online. 
The co-alignment procedure followed the approach used by
\citet{Welsch2012}, described in their Appendix.

By combining $\vvec_h$ estimated with FLCT with $\bvec$ from the SP
magnetogram, \citet{Yeates2014} estimated the average Poynting flux to
be $1.7 \times 10^7$ erg cm$^{-2}$ s$^{-1}$.  This energy flux is
less than the combined energy demand for the chromosphere and corona
in active regions estimated by \citet{Withbroe1977}.
As discussed in greater detail below, however, this estimate did not
account for the observationally estimated magnetic filling factors that
had been applied to each magnetic field component in the vector
magnetogram used by \citet{Schrijver2008}.
Insufficient Poynting flux would indicate that processes on spatial or
temporal scales that are unresolved in these photospheric observations
(e.g., waves or smaller-scale footpoint shuffling) play a significant
role in heating.

Despite the central role of the Poynting flux in theories of coronal
heating, very few observational estimates of Poynting flux in the
context of coronal heating have been published. \citet{Tan2007}
investigated a ``proxy Poynting flux,'' $|\vvec_h| B_{\rm LOS}^2$ in
more than 160 active regions, determined by applying LCT to LOS
magnetograms, and found values in the range $10^{6.7}$ -- $10^{7.6}$
erg cm$^{-2}$ s$^{-1}$. They used the LOS fields alone because
sequences of vector magnetic field measurements were quite rare.  More
recently, \citet{Kano2014} estimated the work done by flows on the
magnetic field, by parametrizing the expected deformation of coronal
loop structure by surface flows.  The expression they derive is
inversely proportional to coronal loop length $L \sim 100$ Mm, and for
LCT flows with typical magnitudes of 0.5 km s$^{-1}$, they
estimate a Poynting flux of $2 \times 10^6$ erg s$^{−1}$ cm$^{−2}$.
They also found typical flow speeds were lower in regions with higher
magnetic filling factors.  This is roughly an order of magnitude lower
than the value of $1.7 \times 10^7$ erg s$^{−1}$ cm$^{−2}$ reported by
\citet{Yeates2014}.  Given this considerable variation in published
Poynting flux estimates, further investigation of Poynting fluxes is
warranted.

Here, we report additional estimates of the Poynting flux from
plage magnetic fields in the same active region studied by
\citet{Yeates2014}.  Our primary aim is to investigate the
properties of photospheric Poynting flux in greater detail than was
done previously, including its dependence on photospheric magnetic
field structure.
The remainder of this paper is organized as follows.  In the next
section, we briefly describe the magnetic field data and tracking
methods we used to estimate $\vvec_h$.
%
%(More detail is given in \citealt{Welsch2012}.)  
%
In Section \ref{sec:results}, we first present our estimates of the
Poynting flux in another small region of plage in AR 10930, then analyze the
Poynting flux's correlations with magnetic structure in the region.
%, and compare results
%between this region and that studied by \citet{Yeates2014}.  
The region of the NFI FOV that we analyze here was also observed in Ca
II by SOT's Broadband Filter Imager (BFI; Tsuneta \etal \,2008),
\nocite{Tsuneta2008} and in \S \ref{subsec:ca_ii} we compare this
chromospheric emission with the spatial distributions of Poynting flux
and magnetic field components. Finally, we conclude with a brief
discussion of our results in Section \ref{sec:discussion}.

%%%%%%%%%%%%%%%%%%%%%%%%%%%%%%%%%%%%%%%%%%%%%%%%%%%%%%%%%%%%%%%%%%%%%%
\section{Data \& Methods}
\label{sec:data}

\subsection{NFI Magnetograms}
\label{subsec:data_nfi}

Many aspects of the NFI magnetograms that we track to estimate
$\vvec_h$ are described by \citet{Welsch2012}.  These Fe I 6302 \AA
(shuttered) magnetograms of AR 10930 have 0.\arcsec16 pixels, and
were created from the Stokes $V/I$ ratio in Level 0 data.  The data
were recorded between 12-Dec-2006 at 14:00 and 13-Dec-2006 at 02:58,
with a cadence of 121.4 $\pm$ 1.2 s, except for three gaps of 10
minutes and two relatively small time steps of 26 s each.  The
USAF/NOAA Solar Region Summary issued at 24:00 UT on 12-Dec-2006
listed AR 10930 at S06W21, meaning it was relatively near disk center
during the interval we study.
Since the diffraction limit of SOT is near 0.\arcsec32, we rebinned
the NFI magnetograms $(2 \times 2)$.
During this era of the {\em Hinode} mission, a bubble present within
the NFI instrument degraded image quality in the upper part of the NFI
field of view; we ignore pixels from this region in our analyses.  

We converted the measured Stokes $I$ and $V$ signals into
pixel-averaged flux densities, which we denote $\bnfi$, using the
approximate calibration employed by \citet{Isobe2007}.  While the
linear scaling in this approach breaks down in umbrae, it should not
be problematic for plage regions.  Note that we use evolution in image
structure in the NFI magnetograms to derive velocities, but do not use
the estimated flux densities directly in any calculations; for
correlation tracking, what matters is that the images capture the {\em
  spatial structure} of magnetic fields at each time in the sequence.
\citet{Welsch2012} estimated the NFI noise level following
\citet{Hagenaar1999}, by fitting the core of the distribution of flux
densities ($\pm 10$ Mx cm$^{-2}$) in each frame with a Gaussian. Based
upon these fits, they adopted a uniform uncertainty estimate of $\pm$
15 Mx cm$^{-2}$ for $\bnfi$ over the 13-hour run.

Prior to tracking these magnetograms, \citet{Welsch2012} co-aligned
them in time to remove spacecraft jitter and jumps from pointing
changes.  Spectral analysis showed some power at the orbital
frequency, but no clear evidence of helioseismic p-mode leakage into the
estimated magnetic flux densities.

\subsection{SP Vector Magnetograms}
\label{subsec:data_sp}

As mentioned above, \citet{Schrijver2008} used SP data to estimate the
vector magnetic field in AR 10930, and one of the two vector
magnetograms they analyzed falls within our tracking interval.  The
full SP scan ran from 20:30 - 21:33, with $\sim$ 0.\arcsec3 \, pixels.
From these observations, LOS and transverse magnetic field strengths,
azimuth
% (including an initial ambiguity resolution), 
and fill fraction were determined at each SP slit position, as
described by \citet{Schrijver2008}.  The data were then interpolated
onto a uniform grid in the plane-of-the-sky, with 0.\arcsec32 square
pixels, multiplied by the fill fraction, and annealed to set the
ambiguity resolution.  Notably, in pixels with weak total
polarization, the fill-fraction was set to 1.0.  We refer to this
plane-of-sky (POS) SP magnetogram as the POSSP magnetogram.  To
produce the vector magnetogram used by both \citet{Schrijver2008} and
\citet{Yeates2014}, the resulting fields were then reprojected to
represent the magnetic field on a Cartesian plane and mapped onto a
grid with a pixel scale of approximately 0.\arcsec63 per pixel.  (This
was done to reduce the array size for computational expediency in
extrapolating coronal fields.)  We refer to this reduced-resolution SP
magnetogram as the RRSP magnetogram.  (The RRSP magnetogram produced
by \citet{Schrijver2008} is online, in FITS format, at
http://www.lmsal.com/$\sim$schryver/NLFFF/; file contents are
described in the FITS header comment field.)

When fill fractions are estimated in the process of inverting
spectropolarimetric data to infer the magnetic field, the form of
equation (\ref{eqn:poynting_shear}) should be modified
\citep{Katsukawa2005, Kano2014} to properly account for the filling
factor, $f(x,y)$, 
\be S_z^{\rm \,plage} \simeq - f (\vvec_h \cdot \bvec_h)B_z/4\pi
~. \label{eqn:poynting_ff} \ee
%  (\ref{eqn:poynting_ff})
%
That is, the product of intrinsic field strengths should be weighted
by one factor of $f$.  Since each magnetic field component, $B_i$, in
both the POSSP and RRSP magnetograms was already weighted by $f$,
using these values in equation (\ref{eqn:poynting_ff}) requires
unweighting by multiplying by $1/f$.  Interpolation of the fill
fraction array $f(x,y)$ in POS coordinates to the RRSP grid introduces
enough inaccuracies into the resulting array that multiplying by $1/f$
results in implausibly large values of magnetic field strengths and
Poynting fluxes in some pixels.  Consequently, we only report results
from the POSSP data here.  Throughout the remainder of the manuscript,
values for magnetic fields given in units of Mx cm$^{-2}$ refer to
pixel-averaged flux densities, i.e., $f$-weighted, while values quoted
in G refer to intrinsic field strengths.

\begin{figure}
 \begin{center}
  \includegraphics[width=8cm]{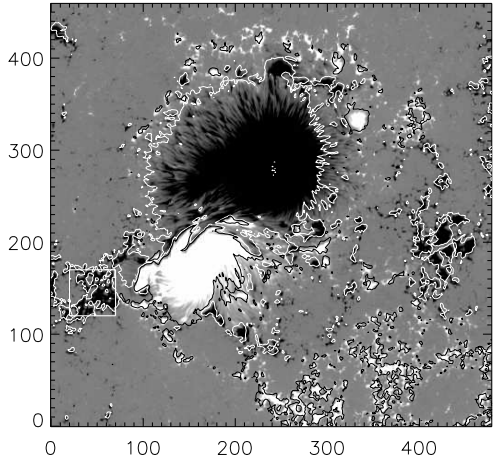} %
 \end{center}
 \caption{Grayscale: $f B_z$ from SP data (saturation at $\pm$
   750 Mx cm$^{-2}$), with $\pm 250$ Mx cm$^{-2}$ contours of $\bnfi$ 
   overplotted (black for flux toward the observer, white for away).
   The plage region that is the focus of our study is in the white
   box at lower left.  Rastering for the SP observation was left-to-right, and
   the observing duration was much longer for the SP data than for the
   NFI data, leading to some local discrepancies between the
   fields.}\label{fig:possp}
\end{figure}

The SP raster across the central part of the active region
that is most closely aligned with the NFI FOV took slightly more than
half an hour.  Since the NFI magnetogram cadence was about two minutes,
% From vxvy_hybrid.sav:
%IDL> print,date_obs_for_most[0]
%12/Dec/06 20:45:34
%IDL> print,date_obs_for_most[478]
%12/Dec/06 21:17:59
no single NFI magnetogram or velocity field is co-temporal with the SP
magnetic field measurements.  Figure \ref{fig:possp} shows $f B_z$ from
the SP data in grayscale, with $\pm 250$ Mx cm$^{-2}$ contours of
$\bnfi$ overplotted (black for flux toward the observer, white for
away).  Rastering for the SP observation was left-to-right, and the longer
SP observing interval causes to some local discrepancies between the
fields.  (In this image, the $x$ coordinates for contours of $\bnfi$
were stretched by 1.01 from the SP data, necessary to compensate for a
small discrepancy $(< 0.\arcsec01)$ found between the NFI and
interpolated SP pixel sizes.)

To analyze approximately simultaneous velocity and magnetic field
data, we restrict our attention to the $(51 \times 51)$ pixel$^2$ area
of plage to the east of the main sunspots in the region.  The plage
region that is the focus of our study in the white box at lower left
of Figure \ref{fig:possp}. The scan across our plage region took about
three minutes, from 20:46:47 -- 20:49:56.  We used the NFI image time
stamped 20:48:20.
%
% FLCT on NFI at 12-Dec-2006 20:48:19.860
%
% cp sp_hires_nfi_overlay.ps Manuscript/fig1.eps
While the SP vector-field estimates are given at plane-of-sky pixel
locations, the magnetic field vectors were expressed in spherical
coordinates $(r, \theta, \phi)$.  We represent these vectors in a
Cartesian coordinate system, with $B_z = B_r$, $B_y = -B_\theta$, and
$B_x = B_\phi$.  While this representation is somewhat inaccurate over
the FOV of the whole active region, it is not problematic in the small
area of plage that we study (about one heliocentric degree on a side).

\begin{figure}
 \begin{center}
  \includegraphics[width=8cm]{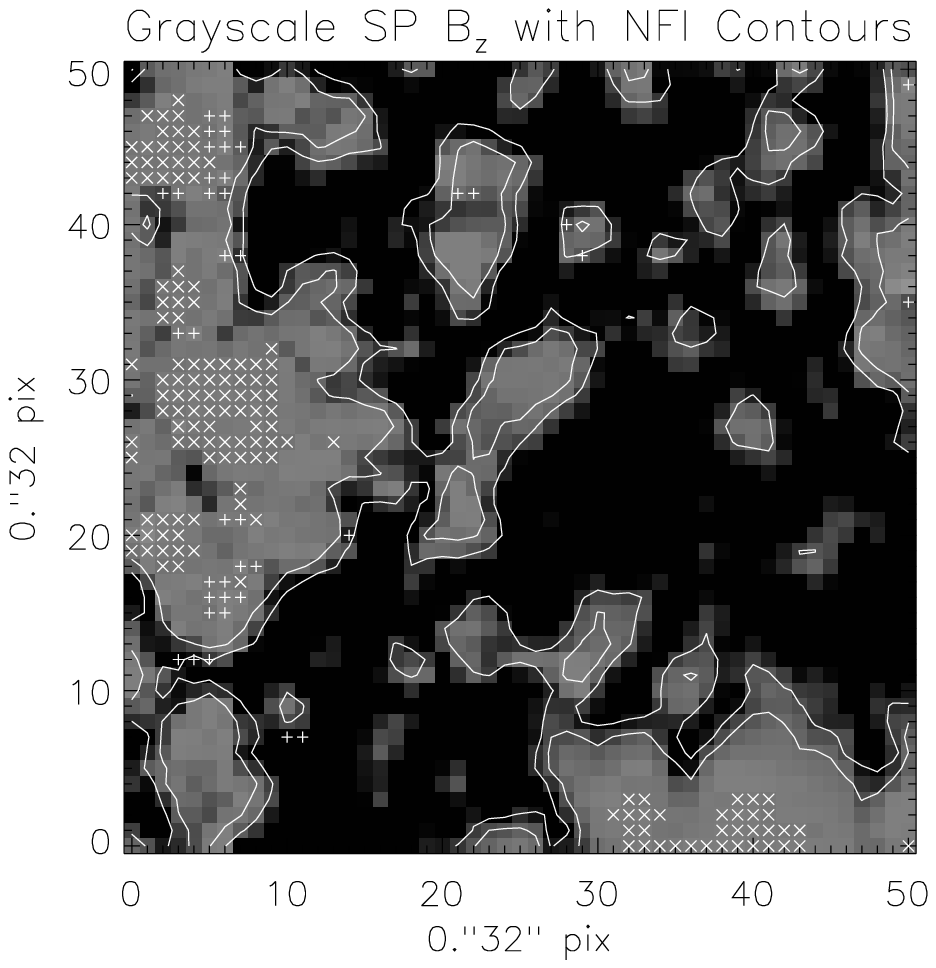} %
  \includegraphics[width=8cm]{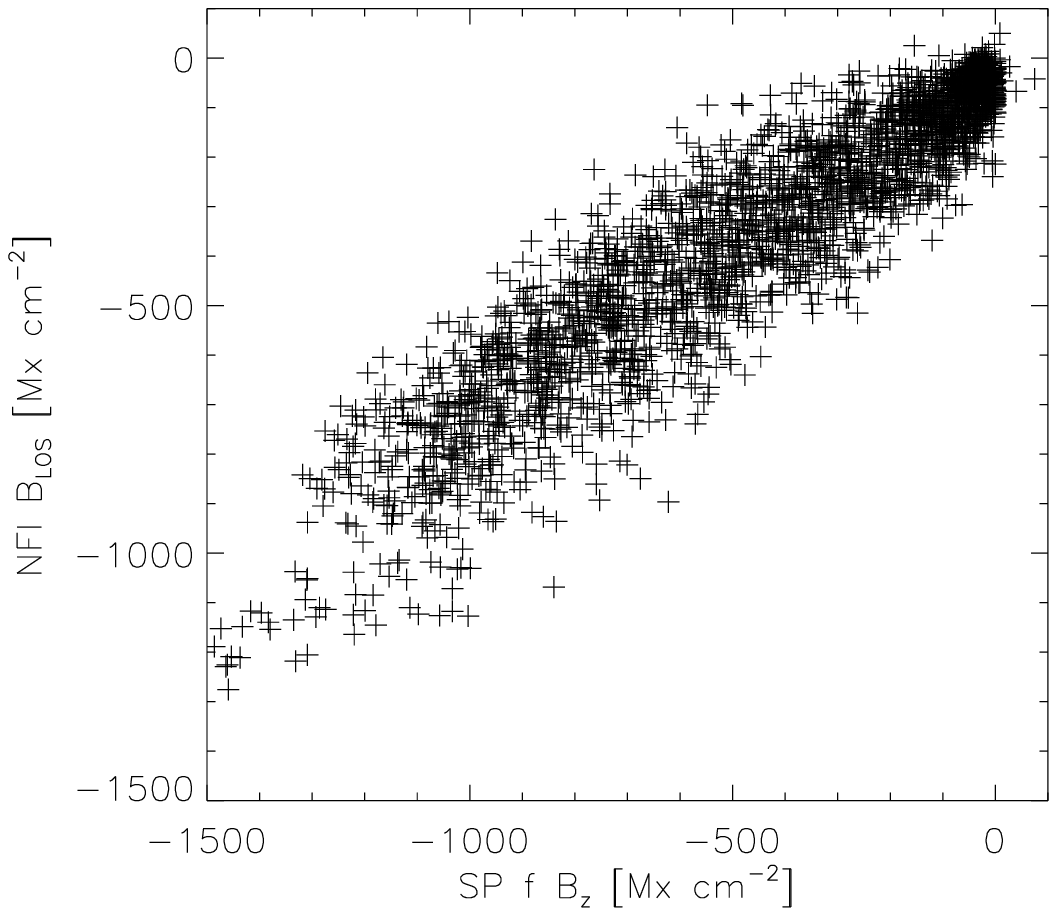} 
 \end{center}
 \caption{Left: Contours of $\bnfi$ at $-125$ and $-250$ Mx cm$^{-2}$
   over a grayscale image of $f B_z$ from the SP data (saturation set
   to $\pm$ 500 Mx cm$^{-2}$) in the plage region in the left panel of
   Figure \ref{fig:possp}.  Errors in $\bvec$ were not estimated in
   pixels marked with $\times$, and were large in pixels marked with
   $+$.  Right: a scatter plot of $f B_z$ from SP data versus $\bnfi$.
   The rank-order and linear correlation coefficients are both 0.93.}
 \label{fig:sp_coreg}
\end{figure}

In the $(2 \times 2)$-binned NFI data, the plage region that we
studied corresponds to $x \in [32,82]$ and $y \in [140,190]$.  We
roughly co-aligned the SP data by hand to within a few pixels, and
then computed the cross-correlation of the nearly-aligned images to
find the whole-pixel shift at the maximum of the cross-correlation
function.  To avoid introducing artifacts from interpolation, we only
co-registered the data down to the pixel scale, and not smaller.
Residual shifts for the SP data in $(x,y)$ are (-0.02,0.23) pixels,
respectively.
% -0.0306950     0.663691 ; metcalf --- old
% -0.0185261    -0.227932 ; new, w/+1
%
To illustrate the co-alignment, we plot contours of $\bnfi$ at $\pm
125$ and $\pm 250$ Mx cm$^{-2}$ over a grayscale image of $f$-weighted
$B_z$ from SP data in the plage region in the left panel of Figure
\ref{fig:sp_coreg}.  In the right panel, we show a scatter plot of
filling-factor-weighted $B_z$ from SP versus $\bnfi$.  The linear and
rank-order correlation coefficients \citep{Press1992} are both 0.93;
the similarity between both measures of correlation implies that
outliers in $\bnfi$ and $B_z$ do not play a major role in the
correlation.
% 0.896574     0.905833 ; metcalf --- old
% 0.932642     0.930758 ; new, w/+1 

A fit of $\bnfi$ to the filling-factor-weighted $B_z$ yields a slope
near 0.69, implying weaker flux densities for $\bnfi$.  Discrepancies
could have arisen from both the more accurate polarimetric measurements
in the SP data and evolution in the fields while the SP was rastered.
% 
% cp lp_sp_hr_nfi_4.ps Manuscript/fig2a.eps
% cp lp_nfi_sp_hr_scatter_4.ps Manuscript/fig2b.eps
%
% +'s are errors too big; x's are no errors / non-inverted
%

The mean and median vertical flux densities in this region are -434 Mx
cm$^{-2}$ and -354 Mx cm$^{-2}$, respectively.  The mean unsigned
$B_z$ has the same magnitude as mean $B_z$, so the region really is
unipolar.  The mean and median horizontal $f$-weighted flux densities
are 160 Mx cm$^{-2}$ and 121 Mx cm$^{-2}$, respectively.  The larger
values of the means compared to medians here imply that some field
strengths are substantially larger than the bulk of the population.
The mean and median inclination angles are similar, 154$^\circ$ and
157$^\circ$, respectively --- so 26$^\circ$ and 23$^\circ$ from
vertical --- implying the field in the bulk of the population is
nearly vertical.

Error estimates for the vector magnetic field were derived by the ASP
inversion code for the SP data in pixels at slit positions with
sufficient polarization signal for reliable inversions, as described
in \cite{Schrijver2008}, and provided by B. Lites (private
communication).  In 139 of our 2601 pixels, error estimates were not
made, corresponding to weak-field pixels; these are shown in the left
panel of Figure \ref{fig:sp_coreg} with $\times$ symbols.
%
%(with mean and median $B_z$ both equal to 32 G).  
%
In 40 pixels, error estimates were large, with errors in inclination
and azimuth exceeding 180$^\circ$ and 360$^\circ$, respectively.
These are also shown in the left panel of Figure \ref{fig:sp_coreg},
but with $+$ symbols.
%
% These also correspond to relatively weak fields, with mean values
% for $(B_x, B_y, B_z)$ of (45, 28, 132) gauss.
%
Among the remaining 2422 pixels, the mean and standard deviation of
uncertainties in field strength, inclination, and azimuth were $(47
\pm 38)$ G, $(2.1 \pm 2.3)^\circ$, and $(7.5 \pm 16)^\circ$.  We
performed simple Monte Carlo simulations to estimate uncertainties in
$(B_x, B_y, B_z)$.  In each run, for pixels with valid uncertainties
(i.e., not the 179 pixels discussed above) we multiplied each pixel's
uncertainties in field strength, inclination, and azimuth by randomly
generated, normally distributed coefficients (appropriately scaled to
the estimated uncertainties), added the results to the original
values, and projected the resulting vector into its $(x, y, z)$
components.  We then computed the mean of absolute differences between
the perturbed and original values for that run.  For 1000 runs, the
uncertainties in $(B_x, B_y, B_z)$ are (30, 36, 35) gauss,
respectively.  The relatively large uncertainty in $B_z$ probably
results from our approach, which averages absolute errors, even though
these might be small in fractional terms for strong-field pixels
$B_z$.  This approach also ignores errors in ambiguity resolutions and
filling factors.

We expect the impact of measurement errors in $\bvec$ on $S_z^{\rm
  \,plage}$ to be relatively small in all summed results: since the
quantities that are summed in equation (\ref{eqn:poynting_ff}) are
signed (from the product of $B_z$ with the dot product of $\bvec_h$
with $\vvec_h$), some cancellation should occur.

% see metcalf_berr.pro

\subsection{Tracking the NFI Magnetograms}
\label{subsec:data_flct}

\citet{Welsch2012} used a tracking code, FLCT \citep{Fisher2008}, to
estimate velocities for the NFI sequence we analyze here.
Many tracking algorithms estimate spatial displacements of local
structures between a pair of images separated in time by an interval
$\Delta t$.  Tracking methods then typically have at least two free
parameters: the time difference $\Delta t$ between images; and the
size $L$ of the local neighborhood (around each pixel for which a
velocity is sought) in which structures between the two images are
associated.  Accordingly, we briefly discuss our tracking parameters.

In some cases, $\Delta t$ is tightly constrained by the cadence of
observations.  If, however, cadences are relatively rapid compared to
the expected time scale of evolution of image structures, then
successive images are likely to differ only by the noise in each
measurement, leading to propagation of noise into the velocity
estimates \citep{Welsch2012}.  \citet{Welsch2012} suggested that
temporal consistency in successive flow maps is a good indicator of
robustness in the velocity estimates.  This can be achieved by
extending $\Delta t$ until significant magnetic evolution has
occurred. Accordingly, the flows we analyze here were derived by
tracking the full NFI FOV with $\Delta t$ = 8 min. Also, the initial
and final magnetograms were computed by applying a five-step boxcar
average to the NFI magnetograms.

Tracking codes (or optical flow methods generally, including LCT,
DAVE, and DAVE4VM; see Schuck 2008)\nocite{Schuck2008} typically
estimate the flow in a given pixel using information about evolution
in a ``local'' neighborhood --- within a user-set length scale, $L$,
that describes the ``apodization window'' or ``aperture'' size ---
around that pixel.  \citet{Schuck2006} noted that, in the presence of
noise, information from several pixels is essential to prevent
spurious fluctuations due to noise from obscuring actual physical
displacements.  Consequently, selecting too small a value for $L$ can
increase susceptibility to noise, since not enough pixels are used in
estimating each local displacement.  Flows smaller than a given scale
$L$ are, however, smoothed over by tracking codes.  We therefore chose
to analyze flow maps derived with $L = 4 \times$ 0.\arcsec32 pixels
(set by FLCT's $\sigma$ parameter, used in a Gaussian windowing
function, $\propto \exp[-r^2/\sigma^2]$), which struck a balance
between boosting correlations between successive flow maps (i.e.,
suggesting the flow estimates were robust) but not over-degrading the
resolution of the magnetograms that were tracked.

We also attempt to minimize confusion of fluctuations due to noise in
the input magnetograms with {\em bona fide} magnetic evolution by not
estimating velocities in pixels below the noise level.  Accordingly,
pixels in the NFI magnetograms with unsigned flux densities below the
15 Mx cm$^{-2}$ noise level estimated by \citet{Welsch2012} were not
tracked.

\begin{figure}
 \begin{center}
  \includegraphics[width=16cm]{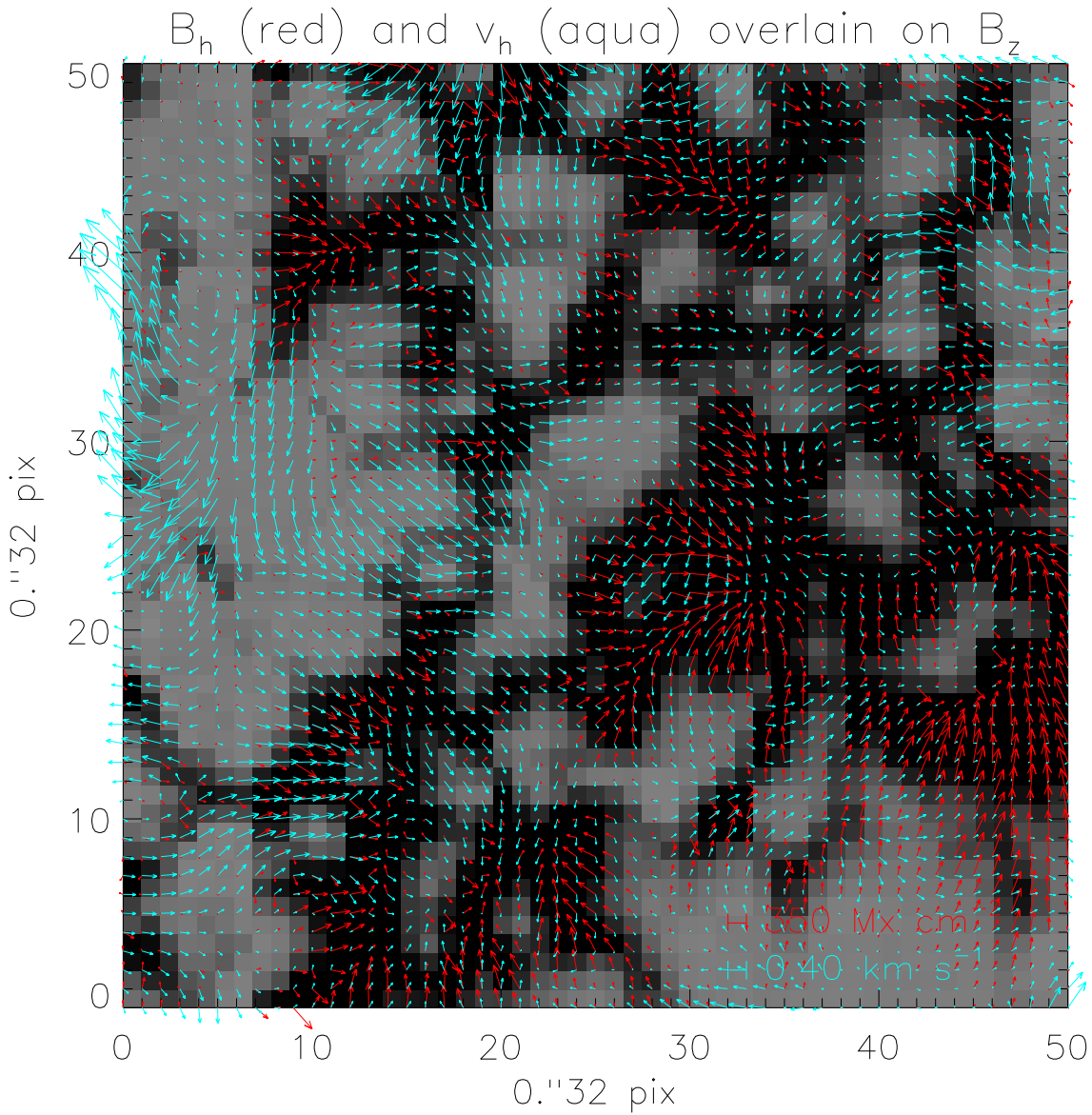} %
 \end{center}
 \caption{Horizontal magnetic field (red) and FLCT velocities (aqua)
   overplotted on a grayscale image of $f B_z$ (saturation at $\pm$
   750 Mx cm$^{-2}$).  These three quantities are necessary
   ingredients for computing the Poynting flux via equation
   (\ref{eqn:poynting_ff}).}
 \label{fig:flct_vectors}
\end{figure}

In Figure \ref{fig:flct_vectors}, we plot both horizontal magnetic
field vectors, $f \bvec_h$, and FLCT flow vectors over a grayscale
image of $f B_z$, for the flow map centered at 20:48:19.
%  
% 
% cp lp_hr_vectors_4.ps Manuscript/fig3.eps ; FF-corrected
The mean and median horizontal FLCT speeds, in the 2,570 pixels in
this field of view where estimates were made, are 0.17 km s$^{-1}$ and
0.14 km s$^{-1}$, respectively.  

Velocities tend to be larger in weaker-field regions, consistent with
the general tendency of strong vertical fields to suppress convection
\citep{Title1992, Berger1998, Bercik2002, Welsch2009, Welsch2012,
  Welsch2013, Kano2014}.  \citet{Kano2014} reported an
anti-correlation between filling factor and the variance in flow
speeds inferred from LCT.  We also found such an anticorrelation in
the plage region studied here: in the 2,414 pixels in which a speed
was estimated and the fill fraction was not 1.0, linear and rank-order
correlations between speeds and fill fractions were both -0.18.  (As
noted in \S \ref{subsec:dependence} below, this value is statistically
significant.)  Anticorrelations were also found between speeds and
each of intrinsic field strength, intrinsic $|B_z|$, and intrinsic
$|\bvec_h|$; but the anticorrelations were weaker than that of speed
with $f$, suggesting that fill fraction is the principal correlated
factor.

As a check upon our results, we also tracked the full NFI FOV with a
separate LCT code, one provided by Y.-J. Moon (private communication)
that has been used in other published work (e.g., Moon \etal
\,2002). \nocite{Moon2002c} While FLCT computes the cross-correlation
function in Fourier space, this second tracking code computes the
correlation function in regular space, following \citet{November1988}.
Hence, we refer to it as Spatial LCT (SLCT), in contrast to Fourier
LCT.  We also only tracked pixels with absolute flux density above 15
Mx cm$^{-2}$, with the same $\Delta t$, but set $\sigma$ in this code
to 3 pixels, since its weighting function includes a factor of 2 in
the denominator of the exponential, $\propto \exp(-r^2/2 \sigma^2)$.
This routine returned either excessively large velocities in some
pixels or even NaNs (in 4\% of tracked pixels).  Velocities in excess
of 2 km s$^{-1}$ ($< 1.5$\% of tracked pixels) or equal to NaN were
set to zero.  

\begin{figure}
 \begin{center}
  \includegraphics[width=8cm]{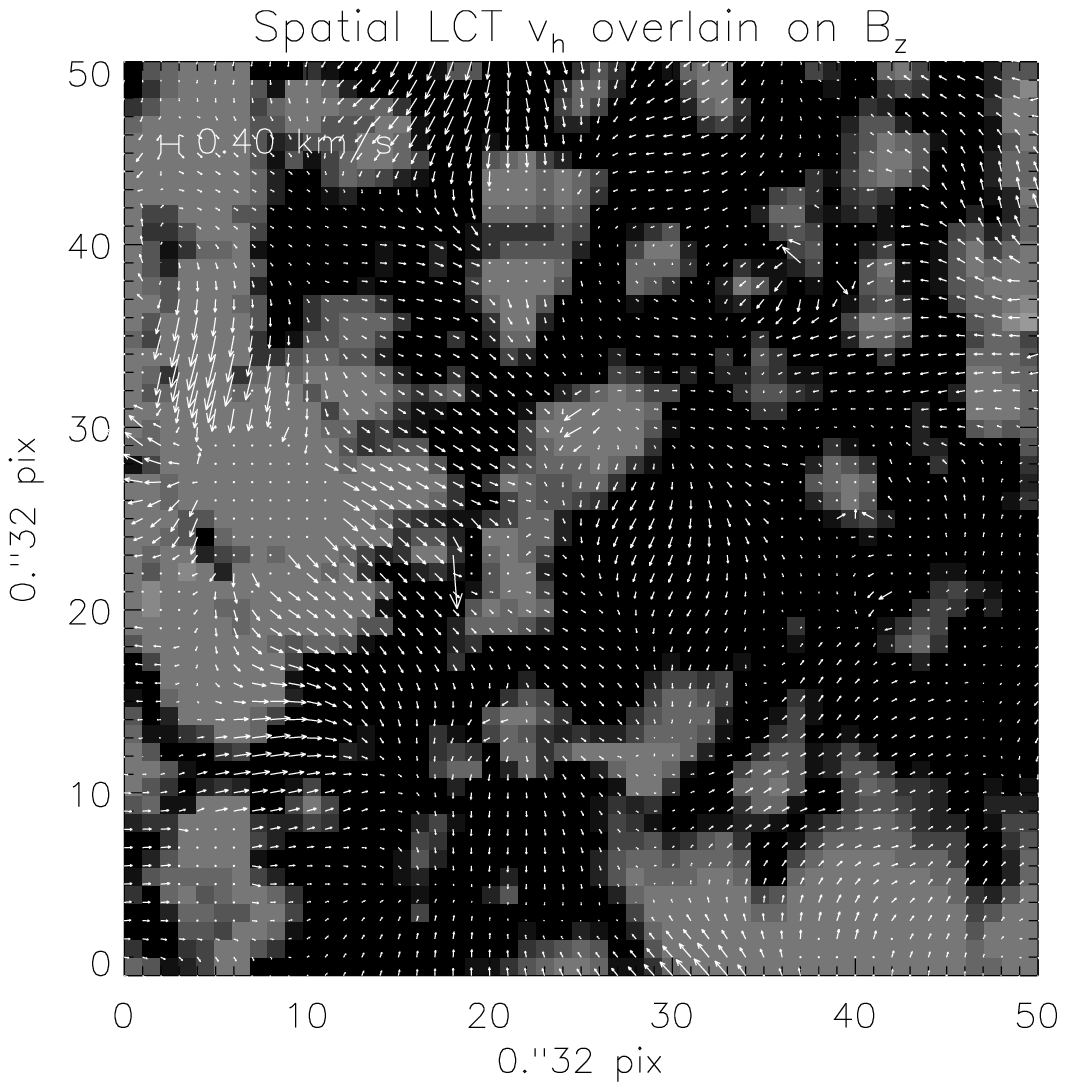} %
  \includegraphics[width=8cm]{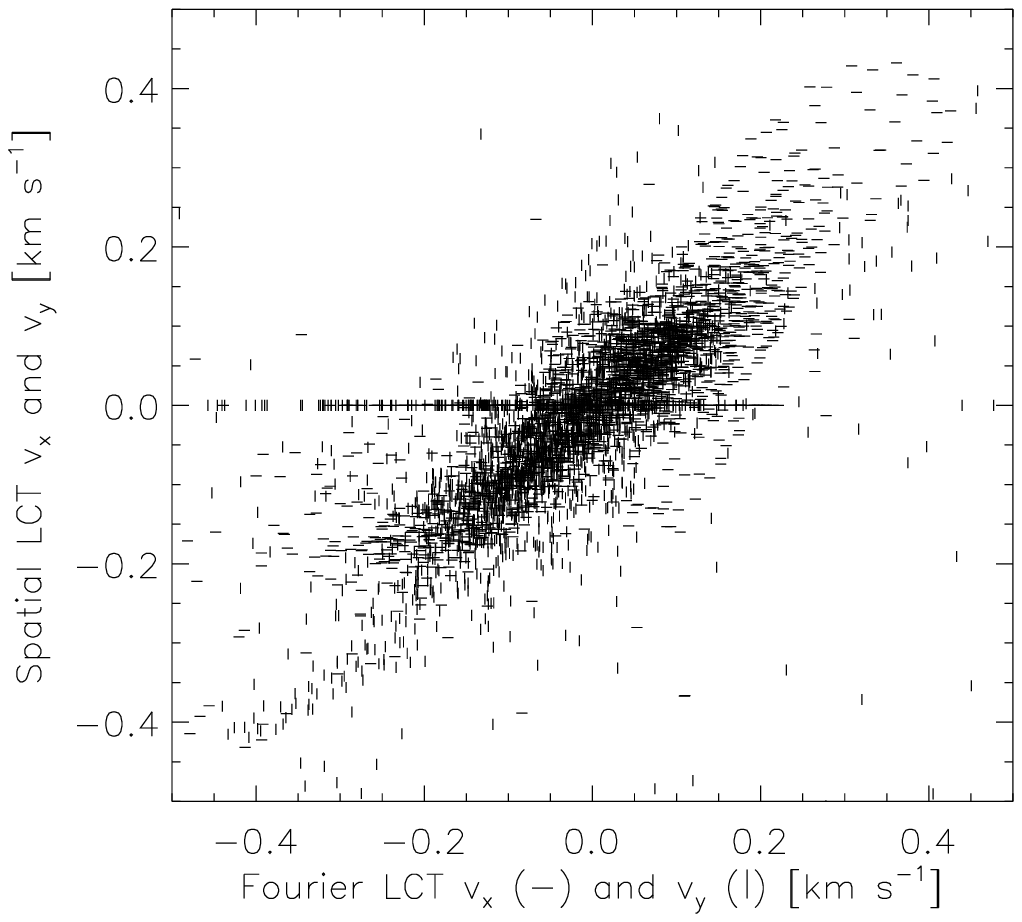}
 \end{center}
 \caption{Left: SLCT velocity vectors overplotted on a grayscale image
   of $f B_z$ (saturation at $\pm$ 500 Mx cm$^{-2}$) from SP.  Right: A scatter
   plot comparing $v_x$ (|'s) and $v_y$ ($|$'s) from FLCT and SLCT.}
 \label{fig:lct_comp}
\end{figure}

In the left panel of Figure \ref{fig:lct_comp}, we show SLCT
velocities overlain on $B_z$.  Comparison of these flows with those in
Figure \ref{fig:flct_vectors} shows rough agreement in many places,
but also clear disagreements in others.  Rank-order correlation
coefficients between these methods' $v_x$ and $v_y$ values in pixels
where both methods made valid estimates were 0.85 and 0.77,
respectively.  Linear correlations were similar, at 0.85 and 0.63 for
$v_x$ and $v_y$, respectively.  Linear and rank-order correlations
near 0.8 result from adding 10\% random variation to a flow component
and then correlating it with the unperturbed flow component.  This
suggests about $\sim$10\% variability in estimated flows due to the
LCT implementation.  Consistent with these significant correlations, a
scatter plot in the right panel of Figure \ref{fig:lct_comp} shows
that the flows are substantially correlated.
% 
% cp moon_vectors_3.ps Manuscript/fig4a.eps
% cp flct_moon_3.ps Manuscript/fig4b.eps
The mean and median horizontal SLCT velocities, among pixels where
valid estimates were made, are 0.15 km s$^{-1}$ and 0.13 km s$^{-1}$,
respectively, quite close to the values for FLCT.

%%%%%%%%%%%%%%%%%%%%%%%%%%%%%%%%%%%%%%%%%%%%%%%%%%%%%%%%%%%%%%%%%%%%%%
\section{Results} 
\label{sec:results}

%%%%%%%%%%%%%%%%%%%%%%%%%%%%%%%%%%%%%%%%%%%%%%%%%%%%%%%%%%%%%%%%%%%%%%
\subsection{Poynting Fluxes}
\label{subsec:poynting}

\begin{figure}
 \begin{center}
  \includegraphics[width=16cm]{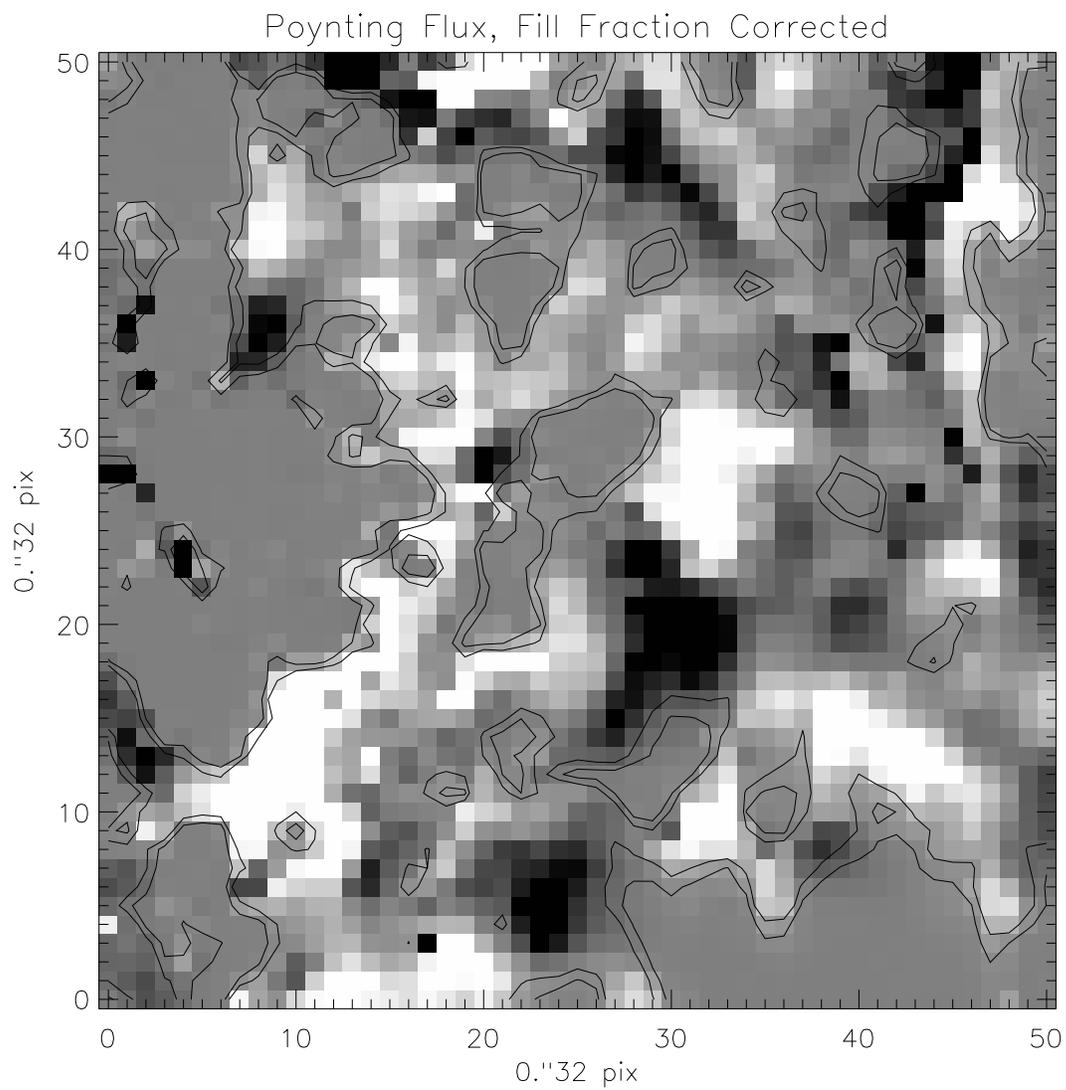} %
 \end{center}
 \caption{The grayscale map shows the Poynting flux, with saturation
   set to $\pm 4 \times 10^{8}$ erg cm$^{-2}$ s$^{-1}$, overlain with
   -125 Mx cm$^{-2}$ and -250 Mx cm$^{-2}$ contours of $f B_z$.
   Regions with both positive and negative Poynting flux are visible,
   but the net Poynting flux is positive.}
 \label{fig:poynt_lp_flct}
\end{figure}

We combined the FLCT flows estimated from the NFI data with the
co-registered vector magnetic field data and fill fraction from SP in equation
(\ref{eqn:poynting_ff}) to compute the Poynting flux averaged over
the plage region.  We find a net positive
average Poynting flux,
$S_z^{\rm \,plage,\,FLCT} = 4.9 \times 10^{7}$ erg cm$^{-2}$ s$^{-1}$ ~. 
%  FLCT: 4.88825e+07
%  SLCT: 5.05869e+07
% 
In Figure \ref{fig:poynt_lp_flct}, we show a grayscale map of the
Poynting flux, with saturation set to $\pm 4 \times 10^{8}$ erg
cm$^{-2}$ s$^{-1}$, overlain with -125 Mx cm$^{-2}$ and -250 Mx cm$^{-2}$ contours of $f B_z$
from the SP data.  Regions with both positive and negative Poynting
flux are visible, but the net Poynting flux is positive.
% 
% cp lp_poynting_ff_4.ps Manuscript/fig5.eps ; FF-corrected
Using the SLCT flows, we also find a net positive Poynting flux, but
estimate $S_z^{\rm \,plage,\,SLCT} = 5.5 \times 10^{7}$ erg cm$^{-2}$
s$^{-1}$.  This is about 12\% larger than the FLCT result.  The
fractional difference between FLCT and SLCT results, compared to their
average, is about 6\%.  The Poynting flux
maps are significantly correlated, with pixel-wise linear and
rank-order correlations of 0.90 and 0.85 in pixels where both methods
made estimates.  
%
%Our simplistic estimate of uncertainties in the
%magnetic field (\S \ref{subsec:data_sp} above) suggests that
%fractional errors in $\bvec$ could be larger, but are probably
%commensurate.  
%
Evidently, the flow estimation process is a source of
at least a $\sim 10$\% uncertainty in our estimates.  Further study of
this same data set, using a different tracking method (e.g., DAVE or
DAVE4VM; Schuck 2006, Schuck 2008) would be worthwhile.
\nocite{Schuck2006, Schuck2008}

Based upon tests of flow reconstruction methods by \citet{Welsch2007},
it is not surprising that different flow methods yield somewhat
different results.
Flows from most of the methods tested by \citet{Welsch2007} were
significantly correlated with both each other and with the true flows.
But flows from the various methods did not agree closely, and most of
the methods only recovered a fraction of the Poynting flux.  Results
about Poynting fluxes from the tests by \citet{Welsch2007}, however,
are probably not applicable here, because the rising-flux-tube
magnetic geometry in the MHD data they used is very different than our
plage region: their field was primarily horizontal, and the Poynting
flux was dominated by the emergence term, not the shearing term.  For
pixels in the upper 95\% of the distribution in $|B_z|$ (the criterion
they used to determine the population they tracked), this can be seen
in a number of statistical measures: the median horizontal field was
five times stronger than the median vertical field; the mean and
median inclination angles (from the vertical) were both larger than
65$^\circ$; and the emergence term in the Poynting flux was largest in
every pixel above their tracking 5\% threshold (and in 99\% of all
pixels).

To characterize the uncertainty in our Poynting flux estimate due to
uncertainties in the magnetic fields, we also employed the Monte Carlo
approach described in \S \ref{subsec:data_sp} above to calculate the
effect of magnetic fields components perturbed by the inversion
uncertainties on the Poynting flux computed via
(\ref{eqn:poynting_ff}).  Excluding the 179 pixels with invalid error
estimates from these Poynting flux calculations, in 1000 runs of
randomly perturbed magnetic fields, we find the mean and standard
deviation of the Poynting flux to be $(5.2 \pm 0.1) \times 10^{7}$ erg
cm$^{-2}$ s$^{-1}$.  If we use the same approach, but substitute the
mean uncertainty estimates from all other pixels for the 179 pixels
with excessive errors, then for 1000 runs we find a mean and standard
deviation of the Poynting flux of $(4.9 \pm 0.1) \times 10^{7}$ erg
cm$^{-2}$ s$^{-1}$.  This suggests that uncertainties in the estimated
magnetic fields are a relatively small part of the overall uncertainty
in the Poynting flux.

\begin{figure}[!htb] %\epsscale{0.75}
\centering
  \includegraphics[width=10cm]{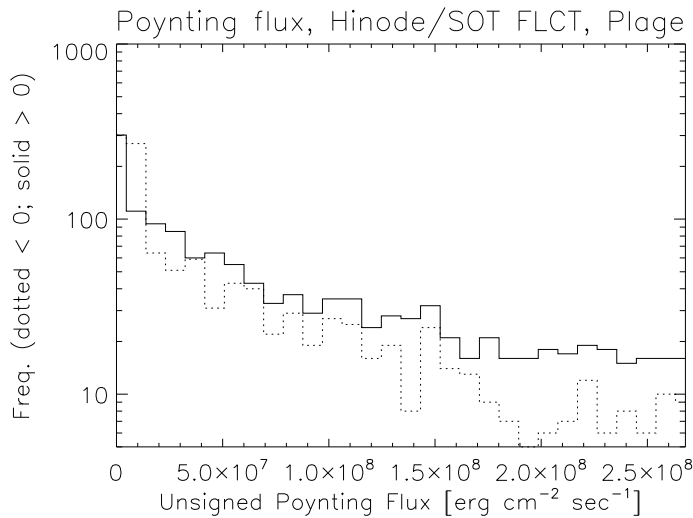} %
\caption{Histograms of the upward (solid) and
    downward (dotted) Poynting fluxes, taken from the map Figure
    \ref{fig:poynt_lp_flct}.  In pixels with high values of unsigned
    Poynting flux, upward-directed fluxes outnumber downward-directed
    fluxes.  This statistical bias suggests that the net-positive
    Poynting flux over the FOV did not arise randomly, due to
    just a few pixels.}
\label{fig:poynt_lp_distrib}
\end{figure}

In Figure \ref{fig:poynt_lp_flct}, many values of the Poynting flux
are much larger than the average value. Could the upward average
energy be an accident, due simply to excess Poynting flux from a few
pixels with large values?  The distribution of Poynting flux values
suggests that the net upward flux arises from a statistical
predominance of upward fluxes in the high-Poynting-flux wings of the
distribution.  This can be seen in Figure \ref{fig:poynt_lp_distrib},
where we plot histograms of the upward (solid) and downward (dotted)
Poynting flux, taken from the map in Figure \ref{fig:poynt_lp_flct}.
As may be seen, there is a prevalence of pixels with upward-directed
Poynting fluxes at high-Poynting-flux values.
%
% cp hinode_distribs_ff.ps Manuscript/fig6.eps ; FF-corrected

The dependence of the mean Poynting flux on the wings of the
distribution implies that the standard deviation of the Poynting flux
values is not a good estimator of the standard error of the mean
(i.e., the uncertainty in our estimate of the mean).  To estimate the
standard error in the mean, we computed $10^4$ bootstrap samples
\citep{Press1992}, which had a mean Poynting flux of $5.0 \times
10^{7}$ erg cm$^{-2}$ s$^{-1}$ and a standard deviation of $5 \times
10^{6}$ erg cm$^{-2}$ s$^{-1}$.  (Note that this mean is over the
subset of pixels with non-zero Poynting flux estimates; the value of
$4.9 \times 10^{7}$ erg cm$^{-2}$ s$^{-1}$ cited above is over the
entire 2601 pixels in our plage region.)  It should be noted, however,
that since FLCT's $\sigma$ was set to 4 pixels, the data are not
strictly independent: the flows are correlated below this scale.  The
magnetic fields also exhibit structure on a similar scale.
Consequently, the assumption of independent data points that underlies
the bootstrap approach is probably violated here, since neighboring
pixels tend to be similar.
%
%Set against this, the plage area that we analyze is much larger than
%this 4 pixel length scale, implying that $(51/4)^2 \simeq 160$
%independent areas are sampled.  But it is nonetheless possible that
%this estimate of the standard error in the mean from population
%sampling is wrong.
%
This test does, however, demonstrate that the mean Poynting flux we
that report does not depend upon values in just a few pixels, because
we found very similar mean values even when resampling the population
of Poynting flux values.

Considering the variation in estimated Poynting flux indicated by the
differing tracking methods ($\sim 6 \times 10^{6}$ erg cm$^{-2}$
s$^{-1}$) and the bootstrap runs (again, $\sim 5 \times 10^{6}$ erg
cm$^{-2}$ s$^{-1}$), we estimate the overall uncertainty level to be
on the order of $1 \times 10^7$ erg cm$^{-2}$ s$^{-1}$.

One aspect of the Poynting flux map in Figure
\ref{fig:poynt_lp_flct} is notable: upward and downward energy fluxes
appear bipolar in some areas (e.g., near pixel coordinates [30,25],
[30,45], and [45,45]). Inspection of the same regions in Figure
\ref{fig:flct_vectors} shows that these bipolar structures arise when
horizontal magnetic fields change direction (e.g., converge) within an
area of horizontal flows that are more uniform on the same spatial
scale.  Qualitatively, this does not accord with the simplistic
picture of braiding of sub-resolution, elemental flux tubes proposed
by \citet{Parker1983}: a substantial Poynting flux is spatially
resolved, and we do not see fluxes winding about each other.  (We
discuss vorticities in both the flow and magnetic fields in \S
\ref{subsec:dependence}, below).

We note that the average Poynting flux that we obtain for this plage
region is substantially larger than the value of $1.7 \times 10^7$ erg
cm$^{-2}$ s$^{-1}$ obtained by \citet{Yeates2014} for a different
plage region.  Inclusion of the $1/f$ factor in our estimate certainly
explains much of the difference.  Without this factor, our average
Poynting flux would be just $2.7 \times 10^7$ erg cm$^{-2}$ s$^{-1}$,
still a a factor of 1.6 larger than that reported by
\citet{Yeates2014}.
% here: -434 Mx cm$^{-2}$ and -354 Mx cm$^{-2}$
% AY14:  365 Mx cm$^{-2}$ and 274 Mx cm$^{-2}$
We note that the mean and median unsigned vertical fields in our plage
region (434 Mx cm$^{-2}$ and 354 Mx cm$^{-2}$, resp.) are larger than the corresponding
values in the region studied by \citet{Yeates2014} (365 Mx cm$^{-2}$ and 274 Mx cm$^{-2}$,
resp.) by factors of $\sim 1.3$.  So differing field strengths might
explain some of the disparity.
% Tot. Poynt. flux, & avg., Moon on AY plage:  2.59791e+25  1.81824e+07

While magnetograms of the full NFI FOV were co-aligned prior to
tracking \citep{Welsch2012}, it is still possible that mean motion of
the plage region we study here, combined with a mean horizontal
magnetic field in the region, could produce the mean Poynting flux we
find.  To investigate this possibility, we computed a region-averaged
Poynting flux, $\bar S_z^{\rm \,plage}$, given by
\be \bar S_z^{\rm \,plage} = - 
\langle \vvec_h \rangle \cdot \langle \bvec_h \rangle 
\langle B_z \rangle /(4 \pi) 
\label{eqn:fov_avg} ~, \ee 
where the angle brackets denote averaging over the $(51 \times 51)$
pixel plage region, and fill-fraction-weighted magnetic field values
(i.e., pixel-averaged values) were used in the averages.  For FLCT and
SLCT flows, we find $\bar S_z^{\rm \,plage,\,FLCT} = -2.7 \times 10^6$
erg cm$^{-2}$ s$^{-1}$ and $\bar S_z^{\rm \,plage,\,SLCT} = -2.3
\times 10^6$ erg cm$^{-2}$ s$^{-1}$.  These region-averaged values are
significantly smaller than and opposite in sign to the net Poynting
fluxes we find above.

%
%bx0 = mean(sp_hr_bx)
%bz0 = mean(sp_hr_bz) 
%by0 = mean(sp_hr_by)
% print,bx0,by0,bz0
%      61.2185      39.4863     -437.497
% print,mean(vx_lp)*bx0*bz0/(4*!pi)*1e+5
% print,mean(vy_lp)*by0*bz0/(4*!pi)*1e+5     
% -3.03099e+06
%  5.68735e+06
% print,-(-3.03+5.69)
%      -2.66

% print,mean(vx_moon_lp)*bx0*bz0/(4*!pi)*1e+5
% print,mean(vy_moon_lp)*by0*bz0/(4*!pi)*1e+5
% -3.55978e+06 ; exclude > 2 km/s (not: -5.39440e+06)
%  5.84030e+06 ; exclude > 2 km/s (not: 6.47402e+06)
% print,-(-3.56+5.84)
%      -2.28

\subsection{Poynting Fluxes in Other Plage Regions}
\label{subsec:other_regions}

Is the systematic prevalence of pixels with upward Poynting flux seen
in Figure \ref{fig:poynt_lp_distrib} a fluke, or is it the norm?  To
settle this question, it would be helpful to analyze the Poynting
flux in other plage regions.  

As noted above, however, the different cadences of the SP raster used
to measure $\bvec$ and the NFI magnetograms tracked to infer
$\vvec_h$ imply that no single velocity measurement is simultaneous
with the vector magnetic field measurement across the region.  So we
cannot simply apply equation (\ref{eqn:poynting_ff}) across the active
region.  This motivated our focus, above, on a relatively small patch of
plage, for which estimates of $\bvec$ and $\vvec_h$ were nearly
simultaneous.

To work around the simultaneity issue, ``rastered'' 2D arrays (in $x$
and $y$) of $v_x(x,y)$ and $v_y(x,y)$ were constructed, by selecting
each column of the 2D array for each velocity component from the
time-slice of the 3D datacubes (in $x, y, t$) of velocities closest in
time to when the SP measurement was made at the corresponding column.
This enables estimating the Poynting flux from the shearing term over
most of the active region (though data near the top of the FOV is
excluded due to the ``bubble'' in NFI).

\begin{figure}[!htb] %\epsscale{0.75}
\centering
  \includegraphics[width=10cm]{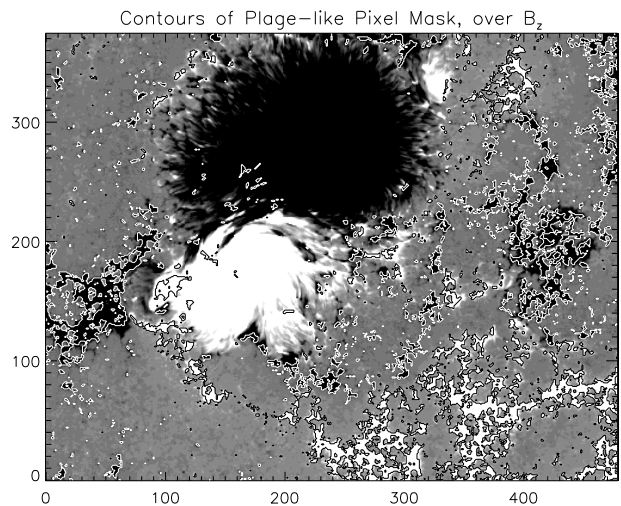} %
\caption{Grayscale shows $f B_z$ over much of AR
    10930, with saturation set to $\pm 250$ Mx cm$^{-2}$.  Contours
    outline regions of ``plage-like'' pixels (white around
    negative-flux regions, black around positive-flux regions), which
    have absolute field strengths between 100 and 1500 Mx cm$^{-2}$
    and inclinations of less than 30 from the vertical, and for which
    fill fractions were estimated (see
    text).}
\label{fig:all_mask}
\end{figure}

Since our focus is on the shearing Poynting flux in plage regions, we
define a mask of ``plage-like'' pixels. We set this 2D bitmap to 1 for
all pixels with filling-factor-weighted $|B_z|$ between 100 and 1500
Mx cm$^{-2}$ and inclinations of less than 30 from the vertical, and
for which fill fractions were estimated.  (Inversions were not
performed for all pixels; fill fractions were not estimated in
non-inverted pixels.  The field in non-inverted regions with
significant Stokes' V signals was assumed vertical.)  Figure
\ref{fig:all_mask} shows $f B_z$ across most of the active region in
grayscale, with contours of the plage-like pixel mask overplotted.
%
% cp all_fov_plage_mask.ps Manuscript/fig7.eps ; FF-corrected
We use the term ``plage-like'' because our criteria for plage
identification are imperfect: a few small regions very near the
positive and negative umbrae satisfy the plage-lake criteria, along
with many very small isolated clumps of quiet-sun fields. Both of
these classes of pixels would probably not be identified as plage by a
human observer.  Our approach does, however, capture the majority of
plage magnetic field regions across the active region.  Further, it is
objective, meaning it can be systematically applied, whereas
identifications made by human observers would be subjective.

\begin{figure}[!htb] %\epsscale{0.75}
\centering
  \includegraphics[width=10cm]{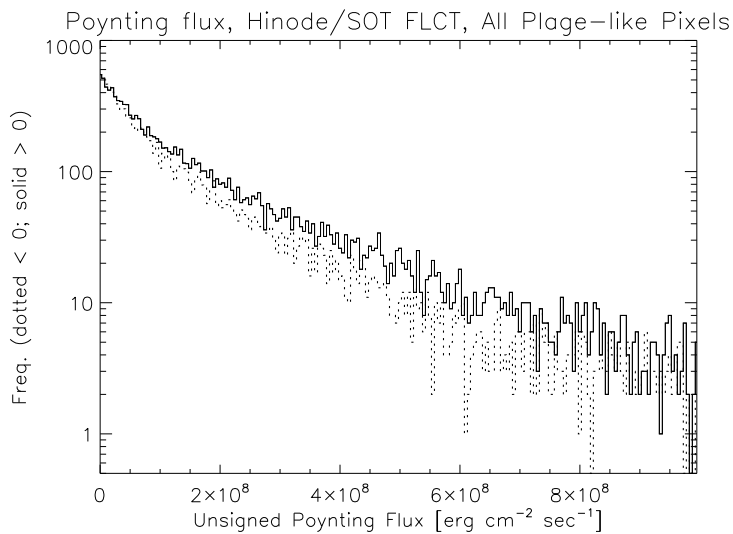} %
\caption{ Histograms of the upward (thin solid) and downward
    (thick dotted) Poynting fluxes from all plage-like pixels (i.e.,
    those within the contours in Figure \ref{fig:all_mask}) that we
    identified in AR 10930.  A systematic bias is evident in the
    frequencies, with upward-flux pixels outnumbering downward-flux
    pixels across most of the distribution.  The systematic
    discrepancy between upward and downward energy fluxes is
    consistent with losses due to atmospheric heating.}
\label{fig:all_distribs}
\end{figure}

Figure \ref{fig:all_distribs} shows the distributions of upward and
downward Poynting fluxes for all plage-like pixels across AR 10930.
As with the distributions from the $(51 \times 51)$ pixel$^2$ region
shown in Figure \ref{fig:poynt_lp_distrib}, the frequency of pixels
with upward Poynting flux is systematically higher than that of pixels
with downward Poynting flux.
%
% cp all_fov_distribs_ff.ps Manuscript/fig8.eps ; FF-corrected
The systematic difference between upward and downward energy fluxes
should correspond to losses of some kind, perhaps due to atmospheric
heating processes.
%Another possibility is the excitation of wave modes, which might leak
%back across the photosphere on shorter temporal and spatial scales
%than we can resolve.

Given the sparse nature of the plage-like pixel mask, and the utility
of closely examining dynamics and magnetic field structure in a sample
plage region, we turn our attention again to the $(51 \times 51)$
pixel$^2$ region that has been our focus.

\subsection{Dependence on Magnetic Structure}
\label{subsec:dependence}

Investigating relationships of magnetic field and flow properties with
the Poynting flux can improve our understanding of the physical
processes that generate Poynting fluxes.  In analogous efforts to
understand coronal heating, \citet{Golub1980} found a clear
relationship between the presence of photospheric magnetic flux and
coronal soft-X-ray (SXR) emission, and \citet{Fisher1998} investigated
relationships between SXR luminosity $L_x$ and global properties of
photospheric magnetic fields several hundred active regions in
Haleakala Stokes Polarimeter vector magnetograms.  Quantities they
analyzed included total unsigned magnetic flux, total unsigned
vertical electric current, and average field strength.
Despite expectations that electric currents should play a role in the
heating that powers coronal SXR emission, they found that the regions'
luminosities depended more strongly on their total unsigned magnetic
flux, $\Phi$, than any other global magnetic variable they
considered. Further, they found that $L_x$ scaled as a power law in
$\Phi$, with an index near one.  \citet{Pevtsov2003} then showed that
the magnetic flux vs. soft X-ray luminosity correlation holds over a
wide range of magnetic scales for the Sun --- from X-ray bright points
to whole active regions to the entire disk --- and even other stars.
(See also Fludra and Ireland [2008], \nocite{Fludra2008} who found
power laws between whole-AR EUV intensities and magnetic fluxes.)

We now apply a similar approach here, but to energy input (the
Poynting flux) as opposed to output (SXR and EUV radiation), and
investigate relationships of magnetic field and flow structure with
the Poynting flux, with the aim of better understanding how Poynting
fluxes arise.  For context when considering other variables, we first
consider baseline correlations between the magnetic field and the
Poynting flux in our $(51 \times 51)$ pixel$^2$ box.  For this, we
only consider correlations for the 2560 pixels (of 2601 total) in
which the velocity was estimated.  Uncertainties in correlation
coefficients can be computed using Fisher's z-transformation, and the
standard error scales like $1/\sqrt{N}$ for correlation coefficients
that are not close to $\pm 1$.  For our sample, the standard error is
about 0.02, so correlations larger than 0.06 in magnitude correspond
to greater than $3\sigma$ departures from the null hypothesis of zero
correlation. Since atypical values for our variables can arise
in our data through errors in the inferred $\bvec$, $\vvec_h$, and
$f$, as well as co-registration, we give rank-order correlation
coefficients, since these are more robust against outliers.

Regarding the {\em unsigned} Poynting flux, we find stronger fields
tend to produce stronger Poynting fluxes.  Correlations between
$|S_z^{\rm \,plage}|$ (from FLCT) and the intrinsic magnetic variables
$|\bvec|, |\bvec_h|$, and $|B_z|$ were 0.65, 0.54, and 0.57,
respectively.  For pixel-averaged values (i.e., $f$-weighted) of the
same variables, the correlations for all three were larger, 0.72,
0.73, and 0.70.  The correlation with fill fraction was positive, at
0.22, suggesting that while flows might be suppressed in pixels with
higher filling factor (as noted above), the stronger fields that tend
to be present lead to larger Poynting fluxes.

What about correlations with the {\em signed} Poynting flux, $S_z^{\rm
  \,plage}$?  Based upon the statistical imbalance in the
distributions of upward versus downward Poynting fluxes, visible in
Figures \ref{fig:poynt_lp_distrib} and \ref{fig:all_distribs}, any
variables correlated with $|S_z^{\rm \,plage}|$ (which is the parent
distribution of the positive- and negative-Poynting-flux
sub-populations) plausibly also exhibit some correlation with
$S_z^{\rm \,plage}$ (the dominant sub-population).
This suggests that sites of larger unsigned Poynting flux should,
statistically, tend have an upward flux, implying the
variables above should also be correlated with $S_z^{\rm \,plage}$,
albeit more weakly than with $|S_z^{\rm \,plage}|$.
Consistent with this idea, we found correlations with the intrinsic
magnetic variables $|\bvec|, |\bvec_h|$, and $|B_z|$ to be 0.12, 0.20,
and 0.09, respectively, while correlations with the corresponding
$f$-weighted variables were 0.18, 0.22, and 0.16, respectively.

Compared to correlations with the magnetic field itself, correlations
with the resolved spatial structure of the magnetic field were weak.
If energy were crossing the photosphere in regions of significant
vertical electric currents, then there should be a strong correlation
between the Poynting flux and the unsigned horizontal curl of the
$f$-weighted horizontal photospheric field ($|\hatz \cdot (\nabla_h
\times \bvec_h)|$, which is $\propto |J_z|$ by Amp\`ere's law).  The
correlation with $|S_z^{\rm \,plage}|$ that we found, however, was
just 0.13 --- while significant, this was much weaker than the
baseline correlations with the magnetic field itself.  The correlation
of $|S_z^{\rm \,plage}|$ with the unsigned, $f$-weighted horizontal
divergence of the horizontal field ($|\nabla_h \cdot \bvec_h|$) was
significantly stronger at 0.28, but also relatively weak compared to
the baseline magnetic correlations.  (This divergence should correspond
to the magnetic field structure in ``azimuth centers,'' albeit for centers
on smaller scales than reported by Martinez-Pillet et al. 1997.)
\nocite{Martinez1997}
Perhaps unsurprisingly, the correlations between {\em signed} Poynting
flux, $S_z^{\rm \,plage}$, and these the unsigned horizontal curl and
divergence of $\bvec$ were even weaker, at 0.04 and 0.07 --- only
marginally significant.

What about correlations between flow properties and the Poynting flux?
First, we found correlations of $|\vvec_h|$ with $|S_z^{\rm \,plage}|$
and $S_z^{\rm \,plage}$ of 0.13 and 0.06, respectively.  The
anti-correlation between filling factor and speed discussed above, and
the dependence of the Poynting flux on the direction of $\vvec_h$
probably both contribute to this weak dependence on speed.

\begin{figure}
 \begin{center}
  \includegraphics[width=15cm]{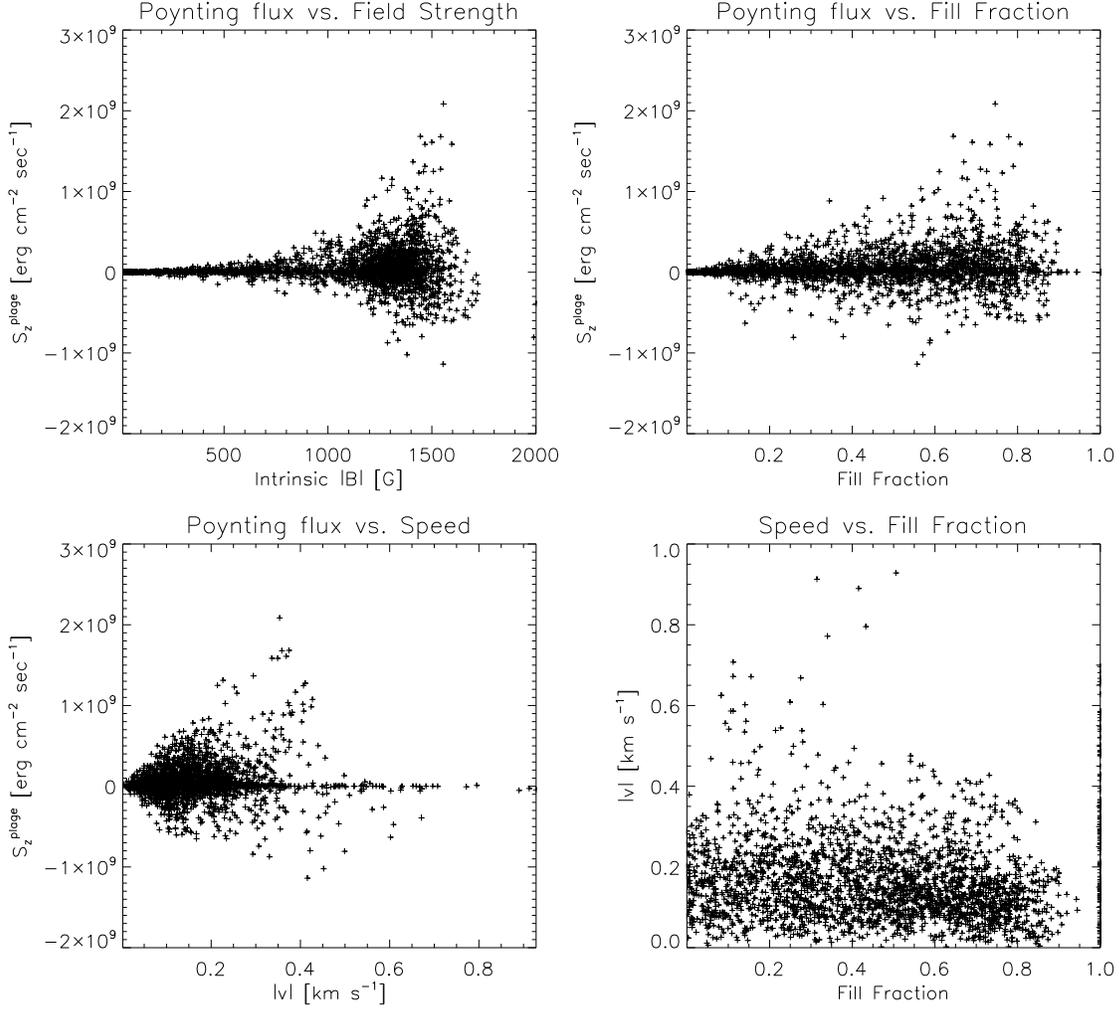} %
 \end{center}
 \caption{Scatter plots of the signed Poynting flux with: intrinsic
   field strength, $|\bvec|$ (upper left); filling factor, $f$ (upper
   right); and speed, $|\vvec|$ (bottom left).  Bottom right:
   scatter plot of speed with fill fraction.}
 \label{fig:scatter}
\end{figure}

We also present some of scatter plots relating some pairs of these
quantities in Figure \ref{fig:scatter}.
The scatter plots of $S_z^{\rm \,plage}$ with intrinsic field strength
(upper left), filling factor (upper right), and speed (bottom left)
show a trend for increasing Poynting fluxes (regardless of sign) as
each of these variables increases.  Also, points with large, positive
values of $S_z^{\rm \,plage}$ tend to outnumber points with large,
negative values of $S_z^{\rm \,plage}$ as each of these variables
increases.
The scatter plot of speed as a function of fill fraction does show a
tendency for higher-speed flows in pixels with lower fill
fractions.
%
% cp scatter_4panel.eps Manuscript/fig9.eps

Vortical motions could play a role in energy transport into the outer
solar atmosphere (e.g., Parker 1983, Kitiashvili et al. 2014),
\nocite{Parker1983, Kitiashvili2014} but we also found relatively weak
correlations between unsigned vorticity, $|\hatz \cdot (\nabla_h
\times \vvec_h)|$, and $|S_z^{\rm \,plage}|$ and $S_z^{\rm \,plage}$
of -0.09 and 0.04, respectively.  Hence, we find little evidence for
resolved vortical flows playing a significant role in driving Poynting
fluxes.  It should be borne in mind, however, that the spatial scale
of resolved by LCT methods is larger than that of the images that are
tracked --- $\sim 1$\arcsec versus 0.\arcsec32 in our case.

It is also possible that converging (or diverging) flows might inject
(or remove) magnetic energy by concentrating (or dispersing) magnetic
flux.  We checked this by correlating $-(\nabla_h \cdot \vvec_h)$,
which should be positive for converging flows, with $|S_z^{\rm
  \,plage}|$ and $S_z^{\rm \,plage}$; both were basically
insignificant at 0.03 and 0.02, respectively.  Correlations with the
unsigned horizontal divergence of $\vvec_h$ were not larger.

It is also worthwhile to characterize the signed energy {\em input}
per unit of magnetic flux, based upon the reported nearly linear
scalings of energy {\em output} in SXR \citep{Fisher1998, Pevtsov2003}
and EUV \citep{Fludra2008} luminosities per unit magnetic flux.
Accordingly, we now compute quantities with units consistent with a
ratio of luminosity per maxwell of $|B_z|$.  For each pixel that was
tracked with FLCT, we computed the ratio of signed energy input per
maxwell. The mean and median of the ratios in this set of pixels were
$1.11 \times 10^5$ erg s$^{-1}$ Mx$^{-1}$ are $6.6 \times 10^4$ erg
s$^{-1}$ Mx$^{-1}$, respectively.
Totaling the energy input and unsigned magnetic flux separately, and
then dividing --- i.e., computing the ratio of sums instead of the sum
of ratios used to compute the mean above --- yields a value of $1.12
\times 10^5$ erg s$^{-1}$ Mx$^{-1}$ for the whole-FOV energy input per
maxwell.
\citet{Pevtsov2003} report SXR luminosities of roughly 10$^3$ erg
s$^{-1}$ Mx$^{-1}$.  Order-of-magnitude estimates of SXR luminosity
$L_x$ as a fraction of total radiated energy from heating $L_{\rm
  heat}$ suggest $L_x \sim 10^{-2} L_{\rm heat}$ \citep{Longcope2004,
  Schwadron2006}.  If the energy fluxes of $\sim 10^5$ erg s$^{-1}$
Mx$^{-1}$ that we find are fully thermalized, and these
order-of-magnitude estimates are correct, then our results are
approximately consistent with those of \citet{Pevtsov2003}.  Studies
of additional plage regions would be worthwhile, to determine if our
value of $\sim 10^5$ erg s$^{-1}$ Mx$^{-1}$ is typical.

%implies a
%trade-off between increasing field strength and decreasing speeds,
%which could mean that very strong fields have weak Poynting fluxes.

\begin{figure}
 \begin{center}
  \includegraphics[width=8cm]{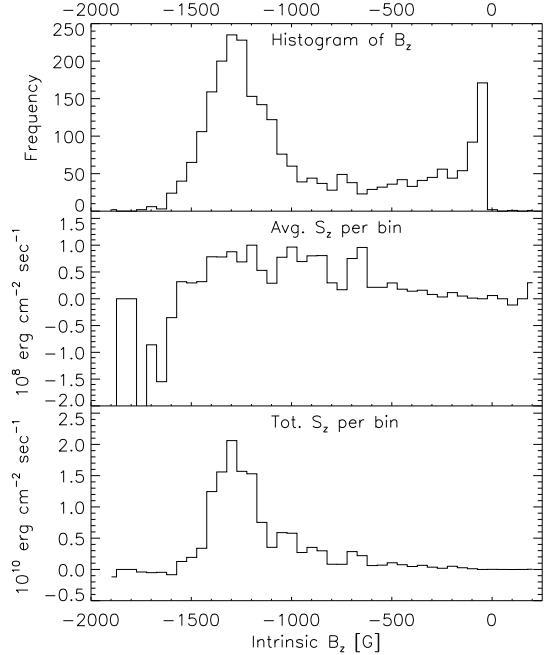} %
 \end{center}
 \caption{Top: A histogram of vertical intrinsic field strengths, with
   clear peak near $|B_z|$ = 1300 G.  Middle: The average of (signed)
   Poynting fluxes over pixels with field strengths in each
   bin. Bottom: The sum of (signed) Poynting fluxes over pixels with
   field strengths in each bin.  The bulk of the contribution to the
   total Poynting flux originates from pixels with vertical field
   strengths near the peak at 1300 G, but pixels with a range of
   vertical field strengths produce, on average, similar Poynting
   fluxes.}
 \label{fig:power_vs_b}
\end{figure}

As we have seen, stronger-field pixels tend to have larger Poynting
fluxes, although in the case of the signed Poynting flux, the
correlation is relatively weak.  The tendency of magnetic fields to
inhibit convection \citep{Title1992, Berger1998, Bercik2002,
  Welsch2009, Welsch2012, Welsch2013, Kano2014} might explain the this
weak correlation: a turning point could be reached as field strength
increases, beyond which increasingly weak convective velocities
produce a smaller convection-driven Poynting flux.  This is a
plausible explanation for the relative darkness of the corona in EUV
and SXR images directly above sunspot umbrae.  (Note, however, that
spatially coherent, large-scale flows, like those in rotating sunspots
[e.g., Brown \etal \,2003], \nocite{Brown2003} could still easily
transport large amounts of magnetic energy across the photosphere in
sunspot fields, though this energy might be more relevant to flares
and CMEs than to coronal heating.)  These considerations raise two
related questions.  First, what is the average (signed) Poynting flux
as a function of field strength?  And second, since some field
strengths are more common than others, which part of the field
strength distribution contributes the bulk of the Poynting flux?  To
address these questions, we first created a histogram of vertical
field strengths, shown in the top panel of Figure
\ref{fig:power_vs_b}.  A clear peak is seen near 1300 G in $|B_z|$.
We then computed the average and total (signed) Poynting fluxes in
each bin (middle and bottom panels, respectively).  From the bottom
panel, it can be seen that the bulk of the total Poynting flux comes
from pixels with vertical field strengths around the peak of the
vertical field strength distribution.  The middle panel shows,
however, that weaker fields, on average, produce a similar Poynting
flux, implying that their smaller contribution to the total energy
flux is due to the relative dearth of such field strengths. Weaker
fields might have average Poynting fluxes as high as stronger fields
because higher velocities tend to be present in the former.
% 
% cp hinode_poynt_bz_ladder.ps Manuscript/fig10.eps
%

%%%%%%%%%%%%%%%%%%%%%%%%%%%%%%%%%%%%%%%%%%%%%%%%%%%%%%%%%%%%%%%%%%%%%%
\subsection{Comparison with Chromospheric Emission}
\label{subsec:ca_ii}

% NFI B_los step 198 is near 20:48
% BFI for AY plage would be step 202, > 21:00
% BFI for left plage should be c. 20:48

It is plausible that regions of enhanced magnetic energy flux across
the photosphere would be brighter in some form of emission. (It is
also possible that the solar atmosphere above the photosphere could
store injected magnetic energy, in the form of electric currents, for
some time prior to its dissipation and consequent enhancement of
emission.  Another possibility is that the energy might propagate away
from the site of its introduction, to be dissipated elsewhere.)

\begin{figure}
 \begin{center}
  \includegraphics[width=8cm]{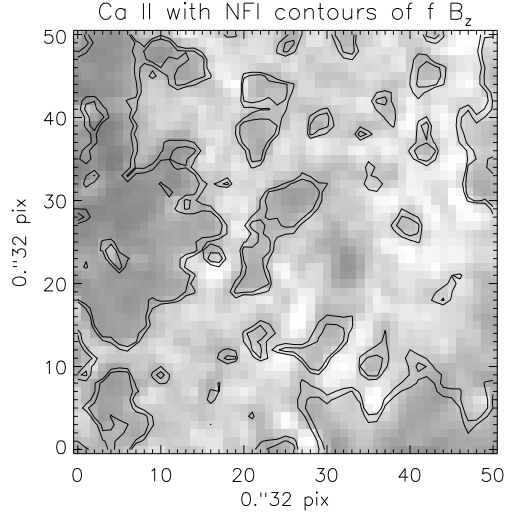} %
 \end{center}
 \caption{$\pm$ 125 and $\pm$ 250 Mx cm$^{-2}$ contours of
   $f$-weighted $B_z$ from SP density are overlain on downsampled Ca
   II intensity from BFI over our plage region.}
 \label{fig:ca_overlay}
\end{figure}

The plage region we analyze here was also observed in Ca II (H line)
by the BFI, so we briefly investigate correlations of the
Poynting flux map and other photospheric magnetic variables with this
emission.  The closest image in time to the NFI velocity estimate was
recorded at 20:48:16 UT on 2006/12/12, which we co-registered with
$\bnfi$ after downsampling from the BFI pixel size of 0.\arcsec11 \,
by a factor of three to approximately match the 0.\arcsec32 scale of
our magnetic field and velocity arrays.  In Figure
\ref{fig:ca_overlay}, we show $\pm$ 125 and $\pm$ 250 Mx cm$^{-2}$
contours of $f$-weighted $B_z$ flux density overlain on the Ca II
intensity in our ($51 \times 51$)-pixel$^2$ plage region.
% 
% cp ca_and_bz.ps Manuscript/fig11.eps

We now investigate correlations of Ca II intensity with magnetic and
velocity field properties, including the Poynting flux, to discover
any interesting relationships.  For magnetic variables, the spatial
map of Ca II intensity exhibited the strongest correlations with
$|B_z|$ and $|\bvec|$, which were greater than 0.6 for both intrinsic
and pixel-averaged field strengths.  The correlation with $|\bvec_h|$
was significantly weaker, at 0.22 and 0.46 for intrinsic and
pixel-averaged field strengths, respectively. Correlations with the
horizontal curl of $f$-weighted $\bvec_h$, $|\hatz \cdot (\nabla_h
\times \bvec_h)|$, and its horizontal divergence, $|\nabla_h \cdot
\bvec_h|$, were statistically significant but much weaker, at 0.14 and
0.28.  The lack of correlation with the curl suggests that electric
current densities do not play a strong role in Ca II emission.  We
also found a significant correlation, 0.42, between Ca II and filling
factor in the subset of pixels in which filling factors were
estimated.

For the Poynting flux, the correlation of Ca II intensity with
unsigned and signed $f$-weighted Poynting fluxes was 0.53, and 0.19,
respectively.  The fact that the magnetic variables that enter the
Poynting flux are more strongly correlated with Ca II emission than
the Poynting flux suggests that the only additional information in the
Poynting flux, from the FLCT flows, is unrelated to Ca II emission;
and, indeed, the correlation of Ca II with $|\vvec_h|$ is
significantly negative, at -0.11.
% added Feb. 20, 15:48
This anticorrelation probably arises because flow speeds are higher in
weak-field regions, while the emission is brightest in strong-field
regions.
Correlations with the horizontal curl of $\vvec_h$, $|\hatz \cdot
(\nabla_h \times \vvec_h)|$, and its horizontal divergence, $|\nabla_h
\cdot \vvec_h|$, were also negative, but marginally insignificant at
-0.07 and -0.08, respectively.  
We also checked the {\em signed} horizontal flow divergence, reasoning
that converging flows might slightly compress the plasma and lead to
heating.  The correlation was also negative (i.e., converging motions
are present slightly more often near brighter Ca II emission), and
stronger but still weak, at -0.12.  This might be related to the
concentration of magnetic flux in downflow lanes where horizontal
flows converge.  
The weak correlation with the curl here implies that resolved braiding
/ vortical motions (e.g., Parker 1983, Kitiashvili et al. 2014)
\nocite{Parker1983, Kitiashvili2014} do
not play a strong role in the generation of Ca II emission.
The reason why the correlation with the curl is negative is hard to
understand, but might be related to the presence of stronger flows in
weaker-field regions.

This analysis suggests that flow information at the spatial and
temporal scales that we study has relatively little bearing on Ca II
emission, compared to magnetic variables.  Tarbell (private
communication) notes that Ca II images from SOT like that which we
analyzed here are ``mostly photospheric'' due to the relatively wide
wavelength band, and suggests that enhanced emission in areas with
strong photospheric fields arises from the hot-wall effect
\citep{Topka1997}, not chromospheric heating.

%%%%%%%%%%%%%%%%%%%%%%%%%%%%%%%%%%%%%%%%%%%%%%%%%%%%%%%%%%%%%%%%%%%%%%
\section{Summary \& Conclusions}
\label{sec:discussion}

By combining LCT velocities estimated from a LOS magnetogram sequence
with a vector magnetogram, both derived from {\em Hinode}/SOT
observations of AR 10930, we estimated the Poynting flux, under the
frozen-in-flux assumption, in a (12 Mm $\times$ 12 Mm) plage region to
be 4.9 -- $5.5 \times 10^7$ erg cm$^{-2}$ s$^{-1}$, depending upon
whether FLCT or SLCT velocities were used.  Errors in the magnetic
fields likely produce smaller uncertainties in the Poynting flux than
this.  These Poynting fluxes are greater than the chromospheric and
coronal energy demands estimated by \citet{Withbroe1977}, $\sim 2
\times 10^7$ erg cm$^{-2}$ s$^{-1}$ and $\sim 1 \times 10^7$ erg
cm$^{-2}$ s$^{-1}$, respectively.

We found that the Poynting flux varied in sign across the plage region
that we studied.  By plotting the distribution of Poynting fluxes in
this region's pixels, we found that the mean upward flux arose from a
predominance of upward-flux pixels toward the high-Poynting-flux end
of the distribution.  We then identified ``plage-like'' pixels ---
those with nearly vertical flux densities in the range 100 -- 1500 Mx
cm$^{-2}$, and sufficient polarization for the vector field to be
estimated --- across the rest of the active region.  The distribution
of Poynting fluxes in this set of plage-like pixels exhibited the same
systematic prevalence of upward-flux pixels, suggesting that the plage
region that is the focus of our study is not a special case.

In analogy with the study by \citet{Fisher1998} relating active
regions' soft X-ray luminosities to magnetic field properties, we
investigated correlations of Poynting fluxes with properties of
the magnetic and velocity fields.  We found that both the {\em
  unsigned} and {\em signed} Poynting fluxes typically increase with
pixels' field strengths.  Correlations of Poynting fluxes with
both unsigned vertical electric current density and flow vorticity
were relatively weak, suggesting braiding or vortical motions (e.g.,
Parker 1983, Kitiashvili et al. 2014) \nocite{Parker1983,
  Kitiashvili2014} are not key aspects of the energy transport
process.  Building upon the work of \citet{Fisher1998},
\citet{Pevtsov2003} found that soft X-ray luminosities for a range of
magnetic regions on the Sun scaled nearly linearly with flux, with a
relationship approximating $\sim 10^3$ erg s$^{-1}$ Mx$^{-1}$.  Here,
we found the energy input per unit magnetic flux to be on the order of
$10^5$ erg s$^{-1}$ Mx$^{-1}$.  We found that fields with intrinsic
vertical field strengths of $\sim 1300$ G supply the bulk of the net
Poynting flux.

We also compared our Poynting flux map with a Ca II intensity image,
and found much stronger correlation of Ca II emission with the
vertical magnetic field strength $|B_z|$ than with vertical Poynting
flux.  We noted that this magnetic correlation might, however, arise
from near-photospheric emission in the passband exhibiting the
hot-wall effect in strong fields.  

The time interval $\Delta t$ between images (eight minutes here) and
windowing length scale $L$ (4 pixels, $\sim 1$ Mm) used in our
tracking will likely filter out processes on shorter temporal and
spatial scales.  Such processes (e.g., waves, or smaller-scale
braiding) might play key roles in chromospheric emission.  Given that
chromospheric and coronal length scales are shorter than the scales we
resolve, observations with higher resolution in space and time (see
below) would be useful to investigate Poynting flux -- emission
correlations further.  It should be noted, however, that our energy
flux is large enough that sub-resolution dynamics are not required to
explain the observed coronal heating.

This initial study leaves several questions unanswered, motivating
related studies to extend the work here.  Do other tracking methods
yield similar results?  The same plage region analyzed here could be
tracked with other methods (e.g., DAVE or DAVE4VM; Schuck 2006, Schuck
2008) \nocite{Schuck2006, Schuck2008} to better understand the
model-dependence of flow estimates in determining Poynting fluxes.
How are photospheric Poynting fluxes related to emission from the
overlying atmosphere?  To address this question, it would be useful to
analyze additional {\em Hinode}/SOT datasets, especially observations
with simultaneous IRIS \citep{DePontieu2014} coverage of
chromospheric, transition region, and coronal emission, to seek any
evidence of spatial or temporal correlations between energy input via
our estimated Poynting fluxes and energy dissipation in the outer
solar atmosphere.  
How rapidly does the spatial distribution of the Poynting flux vary in
time?  In contrast to the snapshot we analyze here, successive
Poynting flux maps would be needed to address this question.  While SP
vector magnetograms are the best currently available, 
%the instrument cannot rapidly raster across large fields of view, and
the telemetry limitations of {\em Hinode} preclude long-duration runs
of successive, rapid rasters over moderately large FOVs.  Consequently, the
HMI instrument aboard SDO \citep{Scherrer2012} could be used
investigate the temporal variation of the Poynting flux.
Unfortunately, HMI has both worse spatial resolution and poorer
spectral sampling.  So a related question is: How sensitively do
estimates of the Poynting flux depend upon a magnetograph's spatial
and spectral resolution?  Analysis of a region simultaneously observed
with SOT and HMI would be worthwhile.
(It is probable, in fact, that additional energy flux could be
resolved with even higher-resolution observations, though the energy
flux must begin to decrease at some limit, to avoid an ultraviolet
catastrophe.  This motivates studies with new, higher-resolution
instruments, as discussed below.)

Within the larger context of the coronal heating problem, we suggest
that a key strategic observational objective for understanding
chromospheric and coronal heating should be construction of a detailed
energy budget for the photosphere-to-corona system, with spatially and
temporally resolved energy inputs correlated with energy release in
all forms --- radiation, kinetic energy in thermal and non-thermal
particles and bulk motion, and gravitational potential energy.  This
will require high-resolution and high-cadence observations of the
magnetic field and emission throughout the photosphere-to-corona
system, for which both space-based observatories (e.g., SDO, IRIS, and
the planned Solar-C
satellite\footnote{http://hinode.nao.ac.jp/SOLAR-C/Documents/Solar-C\_e.pdf})
and existing and planned ground-based observatories (NST [Goode \etal
  2010], GREGOR [Volkmer \etal \, 2010], ATST [Rimmele \etal \, 2010], and
EST [Zuccarello and Zuccarello 2011]) will be essential.
\nocite{Goode2010} \nocite{Volkmer2010} \nocite{Rimmele2010}
\nocite{Zuccarello2011}

%\acknowledgement 
Acknowledgments: We thank the {\em Hinode} science teams for their
hard work in producing the excellent data that made this study
possible. We thank: the referee for constructive criticisms that we
believe improved the manuscript; Anthony Yeates for pursuing
estimation of the Poynting flux in unipolar regions, which inspired
this work; NWRA's K. D. Leka for help with SP data prepared by the
late Tom Metcalf; Bruce Lites for providing error estimates from the
ASP inversions of the SP data and suggestions regarding interpretation
of the data; and T.D. Tarbell and G.H. Fisher for providing useful
comments about the manuscript that helped improve it.  B.T.W. also
gratefully thanks the Japan Society for the Promotion of Science,
whose fellowship supported much of the work that underlies this study,
and acknowledges funding from the NSF's National Space Weather Program
under award AGS-1024862, the NASA Living-With-a-Star TR\&T Program
(grant NNX11AQ56G), and the NASA Heliophysics Theory Program (grant
NNX11AJ65G).  The authors are grateful to Japanese and U.S.  taxpayers
for providing the funds necessary to perform this work.  {\em Hinode}
is a Japanese mission developed and launched by ISAS/JAXA,
collaborating with NAOJ as a domestic partner, and NASA and STFC (UK)
as international partners. Scientific operation of the {\em Hinode}
mission is conducted by the {\em Hinode} science team organized at
ISAS/JAXA. This team mainly consists of scientists from institutes in
the partner countries. Support for the post-launch operation is
provided by JAXA and NAOJ (Japan), STFC (UK), NASA (USA), ESA, and NSC
(Norway).

%%%%%%%%%%%%%%%%%%%%%%%%%%%%%%%%%%%%%%%%%%%%%%%%%%%%%%%%%%%%%%%%%%%%%%
%%%%%%%%%%%%%%%%%%%%%%%%%%%%%%%%%%%%%%%%%%%%%%%%%%%%%%%%%%%%%%%%%%%%%%
%%%%%%%%%%%%%%%%%%%%%%%%%%%%%%%%%%%%%%%%%%%%%%%%%%%%%%%%%%%%%%%%%%%%%%

%\bibliographystyle{/Users/welsch/Library/Latexstuff/apj}
%
%\bibliography{/Users/welsch/Library/Latexstuff/abbrevs,/Users/welsch/Library/Latexstuff/short_abbrevs,/Users/welsch/Library/Latexstuff/full_lib,/Users/welsch/Library/Latexstuff/bib_mods}

\end{document}